\documentclass[aoas,MSNbibl,nameyear,dvips]{arximspdf}
\usepackage{mathbh,shere}
\usepackage{graphicx}

\doi{10.1214/12-AOAS590} 
\volume{7}
\issue{1}
\pubyear{2013}
\firstpage{51}
\lastpage{80}

\makeatletter
\renewcommand{\citep}[1]{(\citeauthor{#1} \citeyear{#1})}
\newcommand{\bs}[1]{\boldsymbol{#1}}
\newcommand{\mc}[1]{\mathcal{#1}}
\newcommand{\mr}[1]{\mathrm{#1}}
\renewcommand{\bm}[1]{\mathbf{#1}}
\newcommand{\ds}[1]{\mathbb{#1}}
\newcommand{\mT}{\mathcal{T}}
\newcommand{\E}{\mathbb{E}}
\newcommand{\V}{\mathrm{\mathbb{V}ar}}
\newcommand{\Note}[1]                              
{\textit{#1}\marginpar[\textbf{$\Longrightarrow$}]
{\textbf{$\Longleftarrow$}}}
\makeatother

\begin{document}
\begin{frontmatter}

\title{Variable selection and sensitivity analysis using dynamic trees, with an application to computer code performance tuning}
\runtitle{Variable selection and sensitivity analysis}

\begin{aug}
\author[A]{\fnms{Robert B.} \snm{Gramacy}\corref{}\thanksref{t1}\ead[label=e1]{rbgramacy@chicagobooth.edu}\ead[label=u1,url]{http://faculty.chicagobooth.edu/}},
\author[A]{\fnms{Matt} \snm{Taddy}\ead[label=e2]{taddy@chicagobooth.edu}\thanksref{t2}}
\and
\author[B]{\fnms{Stefan M.} \snm{Wild}\ead[label=e3]{wild@mcs.anl.gov}\ead[label=u3,url]{http://www.mcs.anl.gov/\textasciitilde wild}\thanksref{t3}}
\thankstext{t1}{Kemper Foundation Faculty Scholar, Econometrics and Statistics group at the University of Chicago
Booth School of Business.}
\thankstext{t2}{Neubauer Family Faculty Fellow, Econometrics and Statistics group at the University of Chicago
Booth School of Business.}
\thankstext{t3}{Supported by the
Office of Advanced Scientific Computing Research, Office of Science,
U.S. Department of Energy, Contract DE-AC02-06CH11357.}
\runauthor{R.~B. Gramacy, M. Taddy and S. M. Wild}
\affiliation{University of Chicago Booth School of Business, University of Chicago
Booth School of Business and Argonne National Laboratory}
\address[A]{R. B. Gramacy\\
M. Taddy\\
University of Chicago Booth School\\
\quad of Business\\
5807 S. Woodlawn Avenue\\
Chicago, Illinois 60637\\
USA\\
\printead{e1}\\
\phantom{E-mail:\ }\printead*{e2}\\
\printead{u1}} 
\address[B]{S. M. Wild\\
Mathematics and Computer Science Division\\
\quad at Argonne National Laboratory\\
9700 S. Cass Avenue, Bldg.~240-1154\\
Argonne, Illinois 60439\\
USA\\
and\\
Computation Institute\\
University of Chicago\\
USA\\
\printead{e3}\\
\printead{u3}}
\end{aug}

\received{\smonth{4} \syear{2012}}
\revised{\smonth{8} \syear{2012}}

\begin{abstract}
We investigate an application in the
automatic tuning of computer codes, an area of research that has
come to prominence alongside the recent rise of distributed
scientific processing and heterogeneity in high-performance computing
environments. Here, the response
function is nonlinear and noisy and may not be smooth or stationary.
Clearly needed are variable selection, decomposition of influence,
and analysis of main and secondary effects for both real-valued and
binary inputs and outputs.  Our contribution is a novel set of tools
for variable selection and sensitivity analysis based on the
recently proposed \textit{dynamic tree} model. We argue that this
approach is uniquely well suited to the demands of our motivating
example.  In illustrations on benchmark data sets, we show that the
new techniques are faster and offer richer feature sets than do similar
approaches in the static tree and computer experiment literature.
We apply the methods in code-tuning optimization,
examination of a cold-cache effect, and detection of
transformation errors.
\end{abstract}


\end{frontmatter}

\section{Introduction}\label{sec1}

The optimization of machine instructions derived from source codes has
long been of interest to compiler designers, processor architects, and
code developers.  Compilers such as \texttt{gcc}, for example, provide a
myriad of flags, each allowing the programmer to choose the ``level''
of optimization.  As codes and their optimization become more complex,
however, it can be harder to know a priori what modifications will
benefit or hinder performance in execution.

Recent advances in the area have demonstrated that higher performance
of a given code can be achieved through annotation scripts (e.g., Orio
[\citet{Orio09}]), which directly apply code transformations such as
loop reordering to the original source to generate a modified, but
semantically equivalent, version of the code.  The output code can
then be compiled in various ways (e.g., by setting compiler
optimization flags or by choosing different compilers), resulting in
an executable that runs on a particular machine more or less quickly
depending on the nature of the transformation, compilation, machine
architecture, and original source code.  Given detailed knowledge of
each aspect of the process, from original source to final executable,
one can obtain significant speedups in execution.  But missteps can
result in significant slowdowns.

Modern high-performance computing
facilities are increasingly complex, making it difficult and/or
time-consuming for a scientific programmer to intimately understand or control
the environment
in which the source code is executed, and thereby affect its performance.
For example, commercial cloud-computing services such as the Amazon Elastic
Compute Cloud (EC2) provide
only a limited description of the available hardware and accompanying
resources;
and scientific and governmental computing
facilities are diverse.  Thus, the need arises to tune codes automatically.

In this paper we report on aspects of a performance-tuning effort
being undertaken at Argonne National Laboratory
to meet needs in scientific computing.
Our aim, given a target machine and source code, is to study how a
suite of
given transformations, together with compiler options
(e.g., \texttt{gcc} flags),
can be used to minimize code execution times under the
constraint that it yields correct output.  As evidenced by the success
of the ATLAS project (\url{http://math-atlas.sourceforge.net/}),
involving a similar but more limited search set, even minor
performance gains for basic computational kernels can be significant
when called repeatedly.


\subsection{A performance-tuning computer experiment}\label{sec1.1}

We focus on data arising from a set of exploratory benchmarking
experiments described by \citet{Balaprakash20112136}.  In the design
of each experiment (the input source code), a subset of the possible
transformation and compilation options (inputs) was thought to yield
correct numerical outputs, and these were varied in full enumeration
over the input space to obtain execution times.  Some of the inputs
are ordinal and some categorical.  Such full enumerations therefore
result in combinatorially huge design spaces---too big to explore in a
time that is reasonable to wait for a compiled executable.  We
investigate the extent to which a statistical model can be used to
measure the relative importance\vadjust{\goodbreak} of each input for predicting execution
times, to explore how each relevant input contributes to the execution
time marginally and (to the extent possible) conditionally, and to
check for any predictable patterns of constraint violations arising
from unsuccessful compilation or runtime errors in the executed code.

Our results in Section~\ref{secapp} show that we can dramatically
reduce the space of options in the search for fast executables: one of
the five inputs is completely removed, and each of the other four has
its range decreased by roughly a factor of $2$.  We perform this
analysis on a dramatically reduced design that, together with a
thrifty inferential technique, means that such information can be
gleaned in an amount of time that most programmers would deem
acceptable to produce a final executable.  We then provide a final
iterative optimization, primed with the results of that analysis, to
obtain a fast executable. The remainder of the fully enumerated
design is then used for validation purposes, wherein we show that our
solution is preferable to alternatives out-of-sample.

Several aspects of this type of data make it unique in the realm
of computer experiments, therefore justifying a noncanonical
approach.  The first is size.  Even when using reduced designs, these
experiments are large by conventional standards.  Second, although
some of the transformation options (i.e., inputs) are ordinal (e.g., a
loop unrolling factor), there is no reason to expect an a priori
smooth or stationary relationship between that input and the response:
for some architectures it may be reasonably smooth, and for others it
may have regime changes due to, for example, being memory-bound versus being
compute-bound.
Third, high-order interactions between the inputs are expected, a
priori, which may prohibit the use of additive models.  Fourth,
checking for valid outputs requires a classification surrogate.
Fifth, since (valid) responses are execution times, the experiment
being modeled is inherently stochastic, whereas many authors define a
computer experiment as one where the response is deterministic.

This last point is perhaps more nuanced than it may seem at first.  In
actuality, many of the sources that can contribute to the
``randomness'' of an executable are known.  For example,
processor loads can be controlled; interruptions from operating system
maintenance threads follow schedules; and
locations in memory, which affect data movement, can be controlled.
But these may more usefully, and practically, be modeled as random.
However, one contributor to the nature of the ``noise'' in the
experiment is of particular interest to the Argonne researchers: the
\textit{cold-cache effect}.

This effect, due to compulsory cache misses sometimes arising from
initial accesses to a cache block, is also referred to as
\textit{cold-start misses} \citet{DPJLbook} and can cause the first
execution instance to run slower than subsequent instances.  It would
be useful to know whether acknowledging and controlling for this
effect are necessary when searching for an optimal executable.
The degree of the cold-cache effect varies greatly from problem to
problem, and determining its significance is vital for designing an
experimental setup (e.g., whether the cache needs to be warmed before
each execution of a code configuration).  Although recent works
[e.g., \citet{PBSWBN11}] have focused on defining input spaces for
performance tuning problems, formulating appropriate objectives in the
presence of the cache effects and other operating system noise remains
an unresolved issue, which application of our techniques can help
inform.  In Section~\ref{secapp} we show that the cold-cache effect
is present but negligible for the particular problem examined.  One
can optimize the executable without acknowledging its effect because
it is very small and does not vary as a function of the input
parameters.\looseness=-1

\subsection{Roadmap}\label{sec1.2}

The remainder of the paper is organized as follows.  Given the unique
demands of our motivating problem set, we make the case in Section~\ref{seccase} that a new, thrifty approach to modeling computer
experiments and decomposing the influence of inputs is needed.  We
maintain that, without using such an analysis to first significantly
narrow the search space, searches for optimal transformations and
compilation settings cannot be performed in a time that is
acceptable to practitioners.  We propose that these needs are
addressed by \textit{dynamic tree} (DT) models, which (along with
previous approaches to model-based decomposition of influence) are
reviewed below and in the \hyperref[app]{Appendix}. In Sections~\ref{secVS} and
\ref{secSA} we improve upon standard methodology for variable
selection and input sensitivity analysis by leveraging the unique
aspects of DTs. Compared with previous tree modeling approaches, our
new methodology offers sequential decision making and fully Bayesian
evidence not previously enjoyed in these contexts.  Compared with the
canonical Gaussian process (GP) model for computer experiments, which
serves as a straw man for many of our comparisons, our methods
facilitate decompositions of input variable influence on problems that
are several orders of magnitude larger than previously possible, while
simultaneously avoiding assumptions of smoothness and stationarity and
allowing for higher-order interactions.  Both sections conclude with
an illustration of the methods, in both classification and regression
applications, and a brief comparison study in support of these
observations.  
Section~\ref{secapp} contains a detailed analysis of our motivating
performance-tuning example using such methods. We conclude in
Section~\ref{secdiscuss} with a brief discussion.

\section{Background}
\label{seccase}

We begin with a review of previous approaches to the analysis of input
influence as relevant to applications in computer experiments,
motivating our dynamic trees approach. These models and accompanying
inferential techniques are then discussed in some detail.

\subsection{Decomposition of influence}

In any regression analysis, one must quantify the influences
on the response by individual candidate explanatory variables.  This
assessment should cover\vadjust{\goodbreak} an array of information, attributing
direction, strength, and evidence to covariate effects, both when
acting independently and when interacting.  For linear statistical
models, various well-known tools are available for the task.  In
ordinary least-squares, for example, there are $t$ and $F$ tests for
the effect of predictor(s), ANOVA to decompose variance contributions,
and  leverages to measure influence in the input space.  Such
tools are fundamental to applied linear regression analysis and are
widely available in modern statistical software packages.

In contrast, analogous techniques for more complicated nonparametric
regression methods, such as neural networks, other basis expansions,
or GPs and other stochastic models, are far less well established.
Many related techniques exist, and we provide a detailed review in our
\hyperref[app]{Appendix}. However, they are not part of the conventional arsenal
applied to the broad engineering problems that motivate this
work---optimization under uncertainty and emulation of noisy computer
simulators---where modeling is further complicated by
nonstationarities manifesting in varying degrees of smoothness.
A~lack of fast, easy-to-apply
tools (and readily available software) means that one typically treats the
response surface model as a black-box prediction machine
and neglects analyses essential for tackling the
application motivating this paper.

To resolve the tension between flexibility and interpretability, we
present a framework that provides both.  We argue that dynamic trees
(DTs), introduced in \citet{taddgrapols2011} and summarized below,
are a uniquely appropriate platform for predictive modeling and
analysis of covariance in complex regression and classification
settings.  Although aspects of DT modeling are just as opaque as, say,
neural networks, they inherit many advantages from the well-understood
features of classic trees.  They take a fast and flexible
divide-and-conquer approach to regression and classification by
fitting piecewise constant, linear models, and multinomial models.
Besides employing a prior that regularizes the nature of the patchwork
fits that result, they make few assumptions about the nature of the
data-generating mechanism.  This approach is in contrast to GP models,
which may disappoint when stationary modeling is inappropriate and
which are burdened by daunting computational hurdles for large
data sets.

Part of our argument holds for tree models in general, of which DTs are
just one modern example: partitioning of the covariate space, the same
quality that is key to model flexibility, acts as an interpretable
foundation for attribution of variable influence.  Distinct from most
other tree methods, however, DTs are accompanied by an efficient
particle sequential Monte Carlo (SMC) method for posterior inference
and can provide full uncertainty quantification for each metric of
covariance analysis, hence allowing for proper consideration of
statistical evidence.  DT inference is also inherently on-line and naturally
suited to the analysis of sequential data.  This aspect is exploited
in our final optimization of the motivating computer experiment.\vadjust{\goodbreak}

Our methodological contributions comprise two complementary
analyses: \textit{variable selection} and input \textit{sensitivity
analysis}.  The first focuses on selecting the subset of covariates
to be included in the model, in that they lead to predictions
of low variance and high accuracy.  The second characterizes
how elements of this subset influence the response.  As discussed in
more detail in the \hyperref[app]{Appendix}, it is most common to focus on only one of
these two analyses: variable selection is common in additive models,
where the structure for covariance is assumed rather than estimated,
whereas in more complicated functional sensitivity analysis settings,
the set of covariates is taken as given.  This methodological split is
unfortunate, because variable selection and sensitivity analysis work best
together, with sensitivities providing a higher-fidelity analysis that
follows in-or-out decisions made after preselecting variables.  Hence,
we have found that the use of DTs as a platform for both analyses is a
powerful tool in applied statistics and is ideal for our motivating
performance-tuning application.

\subsection{Dynamic tree models}
\label{secdynaTree}

The dynamic tree (DT) framework was introduced in
\citet{taddgrapols2011} to provide Bayesian inference for
regression trees that change in time with the arrival of new
observations.  It builds directly on work by
\citeauthor{chipgeormccu1998} (\citeyear{chipgeormccu1998,chipgeormccu2002}), wherein prior models
over the space of various decision trees are first developed.  Since the
Taddy et al.~paper contains a survey of Bayesian tree models and full
explanation of the DT framework, we focus here on communicating an
intuitive understanding of DTs and refer the reader elsewhere for
details.  For those interested in using these techniques, software is
available in the \texttt{dynaTree} [\citet{dynaTree}] package for \textsf{R},
which also includes all the methods of this paper.

\begin{figure}

\includegraphics{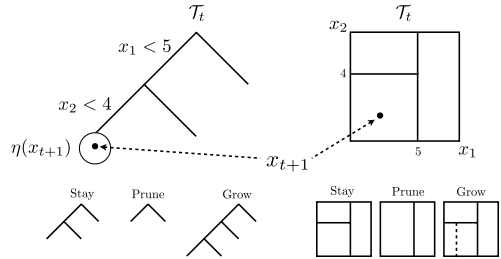}

\caption{Prior possibilities
for tree change $\mT_{t} \rightarrow \mT_{t+1}$ upon arrival of a
new data point at~$\bm{x}_{t+1}$.}\label{treefig}
\end{figure}

Consider covariates $\bm{x}^t =  \{\bm{x}_s\}_{s=1}^t$ paired with
response $\bm{y}^t  =  \{y_s\}_{s=1}^t$, as observed up to time $t$
(the data need not be ordered, but it is helpful to think
sequentially).  A corresponding tree $\mT_t$ consists of a hierarchy
of \textit{nodes} $\eta \in \mT_t$ associated with different disjoint
subsets of $\bm{x}^t$.
This structure is built through a series of
recursive \textit{split rules} on the support of $\bm{x}_t$, as
illustrated in the top row of Figure~\ref{treefig}: the left plot
shows top-down sorting of observations into nodes according to
variable constraints, and the right plot shows the partitioning at the
bottom of the tree implied by such split rules.  These terminal nodes
are called \textit{leaves}, and, in a regression tree, they are associated
with a prediction rule for any new covariate vector.  That is, new
$\bm{x}_{t+1}$ will fall within a single leaf node
$\eta(\bm{x}_{t+1}$), and this provides a distribution for $y_{t+1}$.
For example, a \textit{constant tree} has simple leaf response functions
$\ds{E}[y_{t+1} | \bm{x}_{t+1}] = \mu_{\eta(\bm{x}_{t+1})}$, a {\it
linear tree} fits the plane $y = \alpha_{\eta(\bm{x}_{t+1})} +
\bm{x}^\top\bs{\beta}_{\eta(\bm{x}_{t+1})}$ through the observations in
each leaf, and a \textit{classification tree} uses within-leaf response
proportions as the basis for classification.

Bayesian inference relies on prior and likelihood elements to obtain
a tree posterior, $\mr{p}(\mT_t | [\bm{x}, y]^t) \propto
\mr{p}(y^t|\mT_t, \bm{x}^t) \pi(\mT_t)$.  Given independence across
partitions, tree likelihood is available as the product of likelihoods
for each terminal node; constant and linear leaves use normal additive
error around the mean, while classification trees assume a multinomial
distribution for each leaf's response.  This is combined with a
product of conjugate or reference priors for each leaf node's
parameters to obtain a conditional model for leaves given the tree.
Chipman et al. define the probability of splitting node
$\eta$, with node depth $D_{\eta}$, as $p_{\mathrm{split}}(\mT_t, \eta) =
\alpha(1+D_\eta)^{-\beta}$. Hence, the
full tree prior is $ \pi(\mT_t) \propto
\prod_{\eta \in \mc{I}_{\mT_t}} p_{\mathrm{split}}(\mT_t, \eta)
\prod_{\eta \in L_{\mT_t}} [1-p_{\mathrm{split}}(\mT_t, \eta)]$,
where $\mc{I}_{\mT_t}$ is the set of internal nodes and $L_{\mT_t}$
are the leaves.  They show how a taxonomy of choices of $\alpha$ and
$\beta$ map to prior distributions over trees via their depth.

The DT model of Taddy et al.~adopts this basic framework but combines
it with rules for how a given tree can change upon the observation of
new data.  In particular, $\pi(\mT_{t+1})$ for a new tree is replaced
with $\mr{p}(\mT_{t+1} | \mT_{t}, \bm{x}^{t+1})$, where this
conditional prior is proportional to Chipman et al.'s $\pi(\mT_{t+1})$
but restricted to trees that result from three possible changes
to the neighborhood of the leaf containing $\bm{x}_{t+1}$: {\it
stay} and keep the existing partitions, \textit{prune} and remove the
partition above $\eta(\bm{x}_{t+1})$, or \textit{grow} a new partition by
splitting on this leaf.  This evolution from $\mT_{t}$ to $\mT_{t+1}$
via $\bm{x}_{t+1}$ is illustrated in Figure~\ref{treefig}.  The
original DT paper contains much discussion of tree dynamics, but the
founding idea is that this process leverages the assumed independence
structure of trees to introduce stability in estimation: a~new
observation at $\bm{x}_{t+1}$ will change our beliefs only about the
local area of the tree around $\eta(\bm{x}_{t+1})$.

While the moves from $\mT_{t}$ to $\mT_{t+1}$ are designed to be local
to new observations, inference for these models must account for
global uncertainty about $\mT_{t}$.  This is achieved through use of a
filtering algorithm that follows a general particle learning recipe
set out by \citet{CarvJohaLopePols2009}.  In such methods, the
posterior for $\mT_{t}$ is approximated with a finite sample of
potential tree \textit{particles} $\mT_{t}^{(i)} \in \{\mT_{t}^{(1)}
\cdots \mT_{t}^{(N)}\}$, each of which contain the set of
tree-defining partition rules.\vadjust{\goodbreak}  This posterior is updated to account
for $[\bm{x}_{t+1},y_t]$ by first \textit{resampling} particles
proportional to the predictive probability $p(y_t | \mT_{t}^{(i)},
\bm{x}_{t+1})$ and then \textit{propagating} these particles by sampling
from the conditional posterior $\mr{p}(\mT_{t+1} | \mT_{t}^{(i)},
[\bm{x},y]^{t+1})$ (i.e., drawing from the three moves illustrated in
Figure~\ref{treefig}, proportional to each resulting tree's prior
multiplied by its likelihood).  Hence, tree propagation is local, but
resampling accounts for global uncertainty about tree structure.

Although DTs' inferential mechanics are tailored to sequential
applications, such as sequential design or optimization, they can
also provide a powerful tool for batch analysis.  Since the data
ordering can be arbitrary, it can be helpful to run several
independent repetitions of the SMC method each with a different random
pass through the data.  This approach allows one to study the Monte Carlo error
of the
method, which can be mitigated by averaging inferences across
repetitions.  Such averaging is especially important for Bayes factor
estimation [\citet{taddgrapols2011}].

\section{Variable selection}
\label{secVS}

Tree models engender basic variable selection\break through the estimation
of split locations: any variable not split on has been
deselected.  However, this binary determination does not provide any
spectrum of variable importance, and the unavailable null distribution
for tree splits can lead to inclusion of spurious variables.  Hence,
we need measures of covariate influence that are based on analysis of
response variance. 
Combining these with the full probability model provided by DTs, one
can obtain a probabilistic measure of variable importance and
evidence for inclusion.

\subsection{Measuring the importance of predictors}

Following the basic logic of tree-based variable selection, variables
contribute to reduction in predictive variance through each split
location.  We label the leaf model-dependent uncertainty
reduction for each node $\eta$ as $\Delta(\eta)$.  Grouping these by
variable, we obtain the importance index for each covariate $k \in
\{1,\ldots, p\}$ as
\begin{equation}
\label{eqJ} J_k(\mathcal{T}) = \sum_{\eta \in
\mathcal{I}_{\mathcal{T}}}
\Delta(\eta) \mathbh{1}_{[v(\eta) = k]},
\end{equation}
where $v(\eta) \in \{1,\ldots, p\}$ is the splitting dimension of
$\eta$ and $\mathcal{I}_{\mathcal{T}}$ is the set of all internal
tree nodes (i.e., split locations).  Through efficient storage of data and split rules, these
indices are inexpensive to calculate for any given tree.  Given a
filtered set of trees, as described in Section~\ref{secdynaTree}, the
implied sample of $J_k$ indices provides a full posterior distribution
of importance for each variable; this can form a basis for
model-based selection.

For $\Delta(\eta)$ we consider the decrease in predictive uncertainty
associated with the split in $\eta$.  In regression, the natural
choice is the average reduction in predictive variance,
\begin{equation}
\Delta(\eta) = \int_{A_\eta} \sigma_\eta^2(
\bm{x}) \,d\bm{x} - \int_{A_{\eta_\ell}} \sigma_{\eta_\ell}^2(
\bm{x}) \,d\bm{x} - \int_{A_{\eta_r}} \sigma_{\eta_r}^2(
\bm{x}) \,d\bm{x}, \label{eqredvar}\vadjust{\goodbreak}
\end{equation}
where $\eta_\ell$ and $\eta_r$ are $\eta$'s children,
$\sigma_\eta^2(\bm{x})$ is the predictive variance at $\bm{x}$ in the
node~$\eta$, and $A_\eta$ is the bounding covariate rectangle for that
node.  Rectangles on the boundary of the tree are constrained to the
observed variable support, and, from recursive partitioning, $A_\eta =
A_{\eta_\ell} \cup A_{\eta_r}$.

For constant leaf-node
models, each integral in (\ref{eqredvar}) is simply the area of the
appropriate rectangle multiplied by that node's predictive variance.
For classification, we replace the predictive variance at each node
with the predictive entropy based on $\hat{p}_c$, the posterior
predicted probability of each class $c$ in node~$\eta$.  This leads to
the entropy reduction $\Delta(\eta) = |A_\eta| H_{\eta} - |A_{\eta_\ell}|
H_\ell
- |A_{\eta_r}| H_r$, where $H_\eta = - \sum_c \hat{p}_c \log \hat{p}_c$.
Since the rectangle area calculations involve
high-dimensional recursive partitioning and can be both
computationally expensive and numerically unstable, a Monte Carlo
alternative is to replace $|A_\eta|$ with $n$, the number of data points in
$\eta$.  We find that this provides a fast and accurate approximation.

A regression tree with linear leaves presents a more complex setting,
since the reduction in predictive variance is not constant over each
partition.  In Appendix~\ref{seclinint}, we show that the calculations in
(\ref{eqredvar}) remain available in closed form.  However, since in
this case covariates also affect the response through the linear leaf
model, (\ref{eqJ}) provides only a partial measure of variable
importance.  In Section~\ref{secSA} we describe other
sensitivity metrics whose interpretations do not depend on leaf model
specification.


\subsection{Selecting variables}

An $N$-particle posterior sample
$\{J_k(\mathcal{T}_t^{(i)})_{i=1}^N\}$ can be used to assess the
importance of each predictor $k=1,\ldots, p$, through both graphical
visualization and ranking of summary statistics.  As a basis for
deselecting variables, we advocate estimated relevance probability,
$\mathbb{P}(J_k(\mathcal{T}) > 0) \approx \frac{1}{N} \sum_{i=1}^N
\mathbb{I}_{\{J_k(\mathcal{T}_t^{(i)}) > 0 \}}$.\setcounter{footnote}{3}\footnote{Note that
$\Delta(\eta)$, and thus $J_k(\mathcal{T}_t^{(i)})$ for particle
$i$, may be negative for some $\eta\in
\mathcal{I}_{\mathcal{T}_t^{(i)}}$ because of the uncertainty
inherent in a Monte Carlo posterior sample.}  A backward selection
procedure based on this criterion, illustrated in the examples below,
is to repeatedly refit the trees after deselecting variables whose
relevance probability is less than a certain threshold.

\begin{figure}

\includegraphics{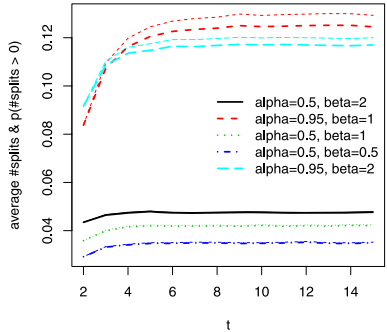}

\caption{Prior splitting frequencies (light) and probabilities of at
least one split (dark) for a 10-dimensional input space plotted by
sample size.  Since all inputs feature equally in the random design,
the results for just one input are shown.}
\label{fprior}
\end{figure}

We use a default relevance threshold of 0.5, such that a variable's
relevance posterior must be less than 50\% negative to entertain
deselection.  However, as we comment in Section~\ref{secccache}, this
can be problematic for some designs, for example, with many categorical
predictors. A more conservative 0.95 threshold has analogy to the
familiar 5\% level for evidence in hypothesis testing, and can be
appropriate in such settings.  For guidance and intuition, one can
refer to the prior distribution on the probability that the tree
splits on a particular input.  Note that this is not the same as a
prior distribution on relevance, which does not exist under our
improper leaf-model priors; rather, the probability of splitting on a
given variable is its probability of having a nonzero relevance.
Figure~\ref{fprior} plots the average number of splits (lighter) and
the probability of at least one split (darker) using the four pairs of
$(\alpha, \beta)$ values explored by \citet{chipgeormccu2002}, plus
the \texttt{dynaTree} default values $(0.95, 2)$, as a function of the
sample size obtained uniformly in $[0,1]^{10}$. These quantities
stabilize after about $t=10$ samples and indicate that, \textit{for this
uniform design}, there is about a 12\% prior probability of
splitting.

Ultimately, the backstop for a proposed deselection is the Bayes
factor of the old (larger) model over the proposed (smaller) one,
terminating the full procedure when proposals longer indicate a strong
preference for the simpler model. Reliable marginal likelihoods are
available through the sequential factorization $p(y^T |\bm{x}^T)
\approx \frac{1}{N} \sum_{i=1}^N \sum_{t=1}^T \log p(y_t | \bm{x}_t,
\mT_{t-1}^{(i)})$ and lead to useful Bayes factor estimates
[see \citet{taddgrapols2011}], as we shall demonstrate.

\subsection{Examples}
\label{secVSexamples}

\subsubsection*{Simple synthetic data}

We consider data first used by \citet{fried1991} to illustrate
multivariate adaptive regression splines (MARS) and then used by
\citet{taddgrapols2011} to demonstrate the competitiveness of DTs
relative to modern (batch) nonparametric models.  The input space is
ten-dimensional, however, the response, given by $10 \sin(\pi x_1 x_2)
+ 20(x_3 - 0.5)^2 + 10x_4 + 5 x_5$ with $\mr{N}(0,1)$ additive error,
depends only on five of the predictors.  Although the true function is
additive in a certain transformation of the inputs, we do not presume
to know that a priori in this illustration. A particle set of
size $N= 10\mbox{,}000$ was used to fit the DT model to $T=$\ 1000 input-output
pairs sampled uniformly in $[0,1]^{10}$.  Following
\citet{taddgrapols2011}, we repeated the process ten times to
understand the nature of the Monte Carlo error on our selection
procedure.

\begin{figure}

\includegraphics{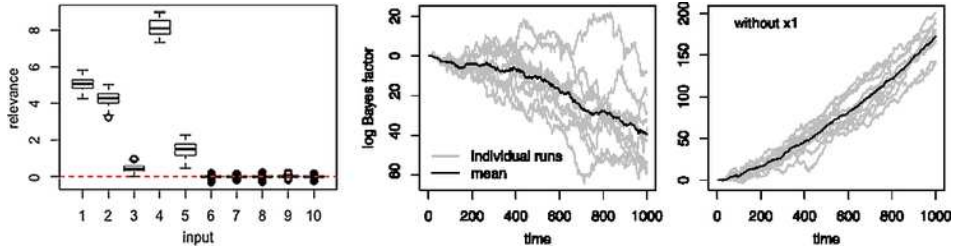}

\caption{Variable selection in the Friedman data.  The boxplots on the
\textit{left} show the posterior relevance.  The \textit{right}
two plots show (log) Bayes factors, first for the full predictor set versus the
set reduced to the five relevant variables, and then
with a relevant variable removed.}
\label{ffried}
\end{figure}

The results are summarized in Figure~\ref{ffried}. The boxplots on
the left show the cumulative 100,000 samples of the tallied relevance
statistics for each variable. The first five all have relevances above
zero with at least 99\% posterior probability. The latter five useless
variables are easily identified, since their relevance statistics
tightly straddle zero. They average about 35\% relevance above zero,
cleanly falling below 95\% or 50\% thresholds. (Figure~\ref{fprior}
is matched to this input domain.) After removing these variables we
reran the fitting procedure and calculated (log) Bayes factors,
treating the smaller model as the null (i.e., in the denominator). All
ten (log) paths (\textit{center} panel) eventually indicate that the
larger model is not supported by the data. In fact, a decreasing trend
in the Bayes factor suggests that the smaller model is actually a
better fit. Thus, while deselecting irrelevant variables is not
technically necessary, doing so becomes increasingly important as the
data length grows relative to a fixed-sized ($N$) particle cloud
(i.e., in order to ward off particle depletion). The right panel in
the figure shows the (log) Bayes factor calculation that would have
resulted if we had further considered the first input for deselection
(i.e., suggesting only inputs 2--5 were important). Clearly, the
larger model (in the numerator) is strongly preferred.


\subsubsection*{Spam data}

\begin{figure}

\includegraphics{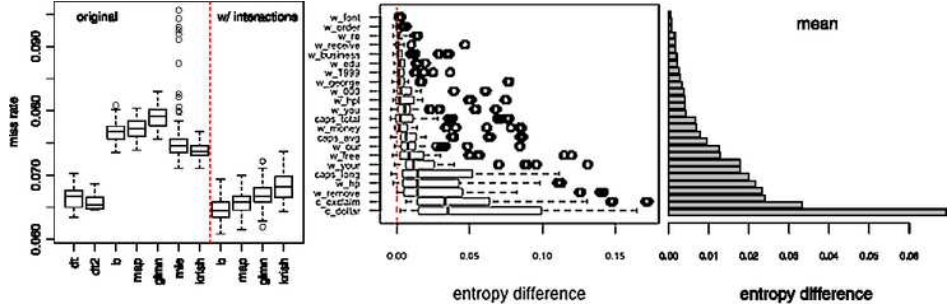}

\caption{(\textit{Left panel}) Boxplots of misclassification rates divided
into two sections, depending on absence or presence of interaction
terms in the design matrix. (\textit{Right panels}) Posterior samples of
relevance statistics and their means.}
\label{fspam}
\end{figure}

We turn now to the Spambase data set from the UCI Machine Learning
Repository [\citet{Asuncion+Newman2007}]. The aim is not only to
illustrate our selection procedure in a classification context but
also to scale up to larger~$n$ and $p$ with significant interaction
effects. The data contains binary classifications of 4601 emails
based on 57 attributes (predictors).
The left panel of Figure~\ref{fspam} shows the results of a Monte
Carlo experiment based on misclassification rates obtained using
random fivefold cross-validation training/testing sets. This was
repeated twenty times for 100 training/testing sets total producing
100 rates. The comparators are modern, regularized logistic regression
models, including fully Bayesian (``\texttt{b}'') and maximum a
posteriori (``\texttt{map}'') estimators via Gibbs sampling
[\citet{grapols2012}], an estimator from the \texttt{glmnet} package [``\texttt{glmn}''; \citet{friedhasttibsh2009}], and
the EM-based method [``\texttt{krish}''; \citet{krishetal2005}].
Results for these comparators on an interaction-expanded set of
approximately 1700 predictors are also provided.  Expansion is crucial
to realize good performance from the logistic models.

Our DT contributions are \texttt{dt} and \texttt{dt2}, each using $N=$\
1000 particles and 30 replicates, which took about half the execution
time of the interaction-expanded logistic comparators. The \texttt{dt2}
estimator is the result of a single iteration of the selection
procedure outlined above using a 50\% threshold (explained below),
leveraging the $\{J_k\}$ obtained from the initial \texttt{dt} run. This
usually resulted in 25 (of 57) deselections. The subsequent Bayes
factor calculation(s) indicated a preference for the small model in
every case considered. Notable results include the following. The
DT-based estimators perform as well as the interaction-expanded linear
model estimators, without explicitly using the expanded predictor set.
Trees benefit from a natural ability to exploit interaction---even a
few three-way interactions were found that, for the other comparators
in the study, would have required an enormous expansion of the
predictor space. Without modification, our new selection procedure
simultaneously allows variables not helpful for main effects or
interactions to be culled. Hence, the estimator obtained after
deselection (\texttt{dt2}) is just as good as the former (\texttt{dt}, using
the entire set of predictors) but with lower Monte Carlo error. In
fact, based on the worst cases in the experiment, \texttt{dt2} is the
best estimator in this study. We found that marginal reductions in
Monte Carlo error can be obtained with further deselection stages.

The right panels of Figure~\ref{fspam} show the posterior samples of
the entropy difference tallies for predictors whose median relevance
was greater than zero; also shown is the corresponding posterior means by
which the samples have been ordered.
A~similar plot is given for random forests in HTF (Figure~10.6). Our
ordering of importance is similar, but importance drops off quickly
because our single-tree model is more parsimonious than are the
additive trees of random forests. As an advantage of our approach, the
middle figure shows posterior uncertainty around these means: there is
a large amount of variability, and evidence of multicollinearity shows
in any given parameter's potential effect ranging from zero to very
large. This observation and an effort to match the size of the
predictor set selected by HTF both contributed to our choice of the
50\% threshold.

\section{Sensitivity analysis}
\label{secSA}

The importance indices of Section~\ref{secVS} provide a
computationally efficient measure of a covariate's first-order
effect---variance reduction directly attributed to splits on that variable.
These indices are not, however, appropriate for all applications of
sensitivity analysis.  First, with nonconstant leaf prediction models,
such as for linear trees, focusing only on variance reduction through
splits ignores potential influence in the leaf model; for example, a
covariate effect that is perfectly linear will lead to $J_k$ near zero
if fit with linear trees. Second, the importance indices depend on
the entire sample and cannot easily be focused on local input regions,
say, for optimization.  Third, the importance indices provide a measure
that is clearly interpretable in the context of tree models but does
not correspond to any of the generic covariance decompositions in
standard input sensitivity analysis.  In this section we describe a
technique for Monte Carlo estimation of these decompositions, referred
to as \textit{sensitivity indices} in the literature, that is model-free
and can be constrained to subsets of the input space.

\subsection{Sensitivity indices}
\label{secST}

The classic paradigm of input sensitivity analysis involves
analysis of response variability in terms of its conditional and
marginal variance.  This occurs in relation to a given
\textit{uncertainty distribution} on the inputs, labeled $U(\bm{x})$.  It
can represent uncertainty about future values of $\bm{x}$ or the
relative amount of research interest in various areas of the input
space [see \citet{TaddLeeGrayGrif2009}].  In applications, $U$ is
commonly set as a uniform distribution over a bounded input region.
Although one can adapt the type of sampling described here
to account for correlated inputs in $U$ [e.g., \citet{SaltTara2002}],
we treat only the standard and computationally convenient independent
specification, $U(\bm{x}) = \prod_{k=1}^p u_k(x_k)$.

The sensitivity index for a set of covariates measures the variance,
with respect to $U$, in conditional expectation given those variables.
For example, the two most commonly reported indices concern
\textit{first-order} and \textit{total} sensitivity:
\begin{equation}
S_j = \frac{\V\{\E\{y | x_j \}\}}{\V\{y\}} \quad\mbox{and}\quad T_j =
\frac{\E\{\V\{y | \bm{x}_{-j}\}\}}{\V\{y\}},\qquad  j =1,\ldots, p, \label{eqinds1}
\end{equation}
respectively.  The first-order index represents response sensitivity
to variable main effects and is closest in spirit to the importance
metrics of Section~\ref{secVS}.  From the identity $\E\{ \V\{ y |
\bm{x}_{-j} \} \} = \V\{y\} - \V \{ \E \{ y | \bm{x}_{-j} \} \}$,
$T_j$ measures residual variance in conditional expectation and thus
represents all influence connected to a given variable.  Hence, $T_j -
S_j$ measures the variability in $y$ due to the interaction between
input $j$ and the other inputs, and a large difference $T_j-S_j$ can trigger
additional local analysis to determine its functional form.  Note that
all moments in (\ref{eqinds1}) are with respect to $U$;  additional modeling
uncertainty about $y |\bm{x}$ is accounted for in posterior simulation
of the indices.

We propose a scheme based on integral approximations presented by
\citet{Salt2002}.  Extra steps are taken to account for an unknown
response surface: ``known'' responses are replaced with 
predicted values. Subsequent integration is repeated across each
tree in a particle representation of the posterior and then averaged
over all particles. Although we focus on first-order and total
sensitivity, full posterior indices for any covariate subset 
are available through analogous adaptation of the appropriate routines
of \citet{Salt2002}.

In the remainder of this section calculations are presented for a given
individual tree; we suppress particle set indexing. Everything is
conditional on a given posterior realization for $y(\bm{x})$.  We
begin to integrate the common $\E^2\{y\}$ terms by recognizing that
\begin{equation}
S_j = \frac{\E\{\E^2\{y|x_j\}\} - \E^2\{y\}}{\V\{y\}}\quad \mbox{and}\quad T_j = 1 -
\frac{\E\{\E^2\{y|x_{-j}\}\} - \E^2\{y\}}{\V\{y\}}. \label{eqinds2}\hspace*{-15pt}
\end{equation}
Assuming uncorrelated inputs, an approximation can be facilitated by
taking two equal-sized random samples with respect to $U$.  Although
any sampling method respecting $U$ may be used, we follow
\citet{TaddLeeGrayGrif2009} and use a Latin hypercube design for the
noncategorical inputs to obtain a cheap space-filling sample on the
margins, thereby reducing the variance of the resulting indices.
Specifically, we create designs $M$ and $M'$ each of size $m$,
assembled as matrices comprising $p$-length row-vectors $\bm{s}_k$ and
$\bm{s}_k'$, for $k=1,\ldots,m$, respectively.  The unconditional
quantities use $M$:
\begin{equation}
\widehat{\E\{y\}} = \frac{1}{m} \sum_{k=1}^m
\E\{y|\bm{s}_k\} \quad\mbox{and}\quad \widehat{\V\{y\}} = \frac{1}{m} \E
\{y|M\}^\top\E\{y|M\} - \widehat{\E^2\{y\}}, \label{eq}\hspace*{-35pt}
\end{equation}
where $\E\{y|M\}$ is the column vector $ [\E\{y|\bm{s}_1\},
\ldots, \E\{y|\bm{s}_m\} ]^\top$ and $\widehat{\E^2\{y\}} =
\widehat{\E\{y\}}\widehat{\E\{y\}}$.

Approximating the remaining components in (\ref{eqinds2}) involves
mixing columns of $M'$ and $M$, which is where the independence
assumption is crucial.  Let $M'_j$ be $M'$ with the $j$th
column replaced by the $j$th column of $M$,
and likewise let $M_j$ be $M$ with the $j$th column of
$M'$.  The conditional second moments are then 
\begin{eqnarray}
\qquad\widehat{\E\bigl\{\E^2\{y|x_j\}
\bigr\}} &=& \frac{1}{m-1}\E\{y|M\}^\top\E\bigl\{y|M'_j
\bigr\},
\nonumber\hspace*{-30pt}
\\[-8pt]
\\[-8pt]
\nonumber
\widehat{\E\bigl\{\E^2\{y|x_{-j}\}\bigr\}} &=&
\frac{1}{m-1}\E\bigl\{y|M'\bigr\}^\top\E
\{y|M_j\} \approx \frac{1}{m-1}\E\{y|M\}^\top\E\bigl
\{y|M'_j\bigr\},\hspace*{-30pt}
\end{eqnarray}
the latter approximation saving us the effort of predicting
at the locations in $M_j$.

In total, the set of input locations requiring evaluation under the
predictive equations is the union of $M$, $M'$, and
$\{M'_j\}_{j=1}^p$.  For designs of size $m$ this is $m(p+2)$
locations for each of $N$ particles.  Together $m$ and $N$ determine
the accuracy of the approximation.  Usually $N$ is fixed by other,
more computationally expensive, particle updating considerations.
Particle-wide application of the above provides a sample from the
posterior distribution for $\bm{S}$ and $\bm{T}$.

\subsection{Visualization of main effects}
\label{secme}

A byproduct of the above procedure is information that can be used to
estimate main effects.  For each particle and input direction $j$, we
apply a simple one-dimensional smoothing of the scatterplot of
$[s_{1j}, \ldots, s_{mj}, s'_{1j}, \ldots, s'_{mj}]$ versus $[\E\{y|M\},
\E\{y|M'\}]$.  This provides a realization of $\E\{y|x_j\}$ over a
grid of $x_j$ values and therefore a draw from the posterior of the
main effect curve.  Note that we use here the posterior means
$\ds{E}[y |\bm{s}]$, as opposed to the posterior realizations for
$y|\bm{s}$ used in calculating sensitivity indices.  Average and
quantile curves from each particle can then be used to visualize the
posterior mean uncertainty for the effect of each input direction as a
function of its value.  One-dimensional curve estimation is
robust to smoother choice in such a large sample size ($2m$); we use a
simple moving average.

\subsection{Examples}
\label{secSAexamples}


Consider again the Friedman data from Section~\ref{secVSexamples},
using the first six inputs.  Ordinarily we would recommend an initial
selection procedure before undertaking further sensitivity analysis to
eliminate all irrelevant variables, but we keep one irrelevant input
for illustrative purposes.

\begin{figure}

\includegraphics{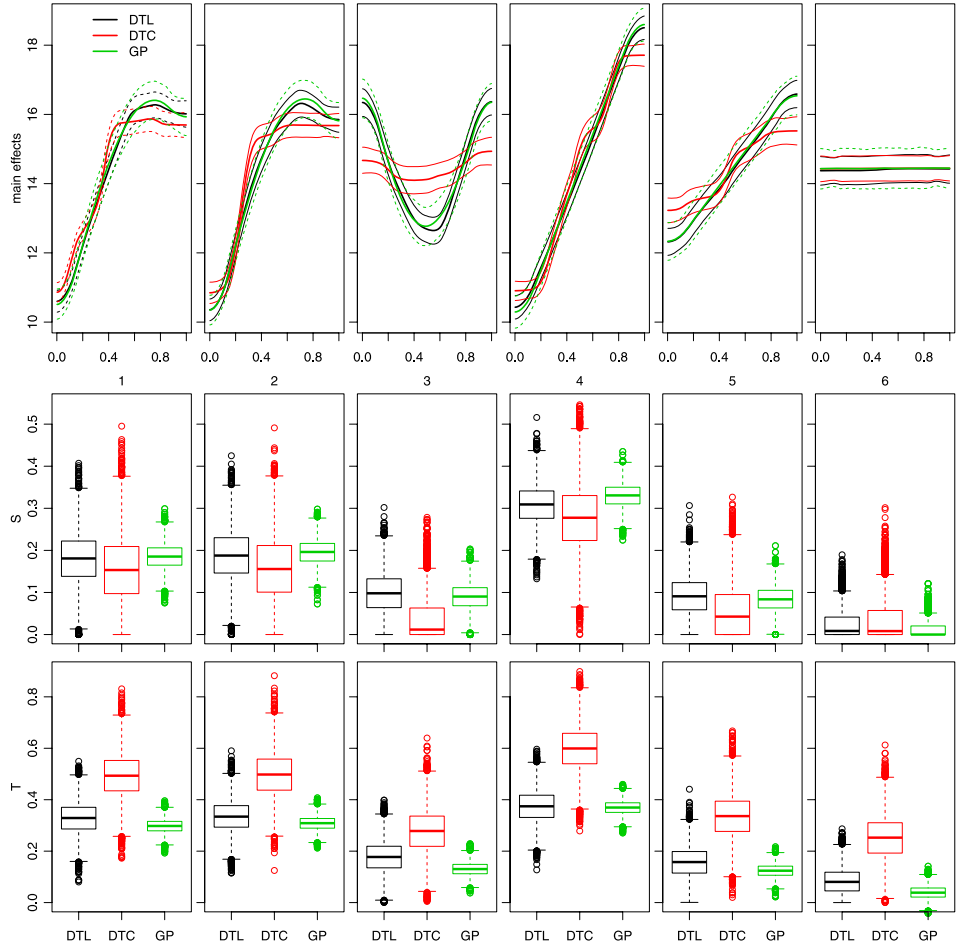}

\caption{Main effects (\textit{first row}) and $\bm{S}$ (\textit{second row}) and
$\bm{T}$ (\textit{third row}) indices for the Friedman data using dynamic
trees and GPs.}
\label{ffriedSA}\vspace*{-6pt}
\end{figure}

Figure~\ref{ffriedSA} summarizes the analysis under constant and
linear DTs (DTC and DTL, resp.), and under a GP (fit using \texttt{tgp}) for comparison.
In all three cases the number of particles
(or MCMC samples for the GP) and samples from $U$ were the same:
$N=10\mbox{,}000$ and $m=1000$, without replicates.  The main effects for DTL
and GP are essentially identical.  As evidenced in the plots, DTC struggles to
capture the marginal behavior of every input; $x_3$ is particularly
off.  These observations carry over to the $\bm{S}$ and $\bm{T}$
indices.  DTL displays the same average values as does the GP, but with
greater uncertainty.  DTC again shows less agreement and greater
uncertainty.  Whereas DTC works well for variable
selection, DTL seems better for decomposing the nature of variable
influence.

With DTL and GP providing such similar sensitivity indices,
why should one bother with DTL?  The answer rests in the
computational expense of the two procedures.  The DT fit and
sensitivity calculation stages each take a few minutes.
The GP version, even using a multithreaded version of \texttt{tgp}, takes
about six hours on the same machine and requires that the two stages
occur simultaneously.  Hence, if new $\bm{x}-y$ pairs are added or a
new $U$ is specified, the entire analysis must be rerun from scratch.
With DTs, the fit can be updated in a matter of seconds, and only the
sensitivity stage must be rerun, leading to even greater savings.  In
sum, the DT analysis can give similar results to GPs but is hundreds
of times faster.\vadjust{\goodbreak}

\begin{figure}

\includegraphics{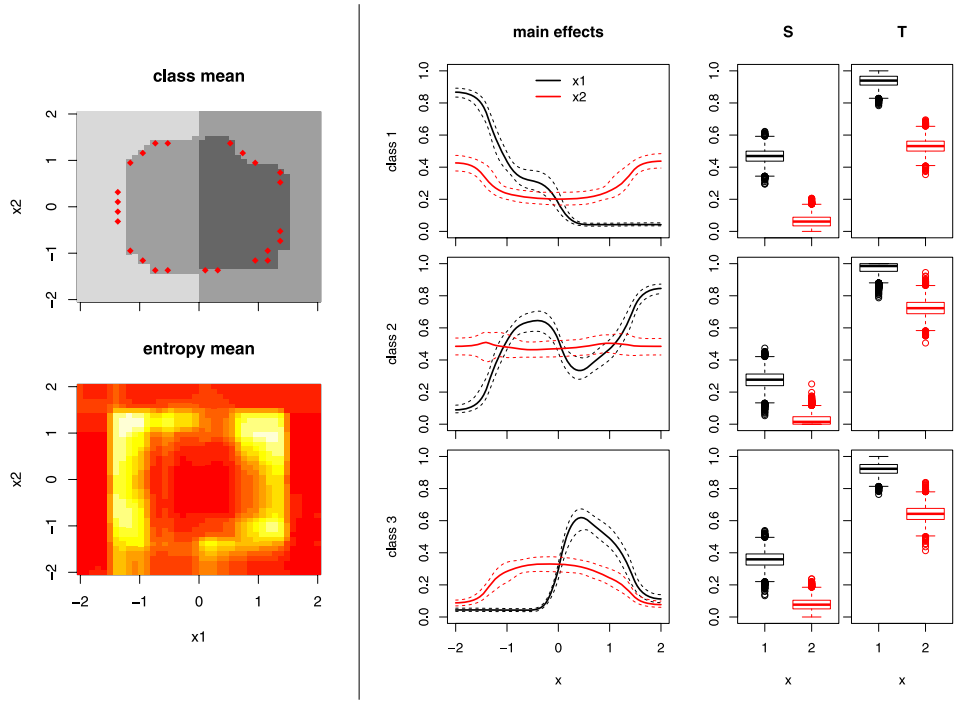}

\caption{(\textit{Left panel}) Posterior predictive mean and entropy;
misclassified points are shown as red dots. (\textit{Right panels})
Sensitivity main effects and $\bm{S}$ and $\bm{T}$ indices for each
class. The black lines and boxplots correspond to input $x_1$, and
the red ones to $x_2$.}
\label{fexpclass}
\end{figure}

GPs also are good (but even slower) at classification (GPC).  Perhaps
this is why we could not find GPC software providing input sensitivity
indices for comparison.  Figure~\ref{fexpclass} shows the results of
a sensitivity analysis for a three-class/2D data set [see
\citet{gramacypolson2011} for details and GPC references].  Fitting a
GPC model from 200 $\bm{x}-y$ pairs takes about an hour, for
example, with the \texttt{plgp} package.  By contrast, fitting a DT with
multinomial leaves using $N=10\mbox{,}000$ particles takes a few seconds; and
the sensitivity postprocessing steps, which must proceed separately
for each class, take a couple of minutes.  The MAP class labels and
predictive entropy shown on the left panel indicate the
nature of the surface.  Notice that the entropy is high near the
misclassified points (red dots).  The smooth transitions are difficult
to capture with axis-aligned splits.

The plots in the right panels show the main effects and $\bm{S}$ and
$\bm{T}$ indices for each class.  All three sets of plots indicate a
dominant $x_1$ influence, which conforms to intuition because that axis
spans three labels whereas $x_2$ spans only two.  Lower $S$ and $T$
values for $x_2$ provide further evidence that its contribution to the
variance is partly coupled with that of $x_1$.

\section{A computer experiment: Optimizing linear algebra kernels}
\label{secapp}

We now examine the data generated by linear algebra kernels from
\citet{Balaprakash20112136}. The execution times for these experiments
were obtained on Fusion, a 320-node cluster at Argonne National
Laboratory.  Each compute node contained a 2.6 GHz Pentium Xeon 8-core
processor with 36 GB of RAM.  We focus here on the GESUMMV experiment.
The results obtained for the other two kernels (MATMUL and TENSOR) we
examined are similar and are therefore omitted because of space
constraints.  GESUMMV, from the updated BLAS library [\citet{UBLAS}],
carries out a sum of dense matrix-vector multiplies.
The tuning design variables considered consist of two loop-unrolling
parameters taking integer values in $\{1,\ldots , 30\}$ and three binary
parameters associated with performing scalar replacement, loop
parallelization, and loop vectorization, respectively.

Argonne allowed an exceptional amount of computing resources to be
assigned to these and a suite of similar performance-tuning examples
in order to study aspects of the tuning apparatus and to enable
initial explorations into elements of the online optimization of
executables such as the one we describe below.  In particular,
resources were allocated for transformation, compilation, and obtaining the
timings of 35 repeated (on the same dedicated node) execution trials at each
design point in a full enumeration of the GESUMMV design space. These
tests incurred over 30 CPU-hours (roughly half of which were devoted
to transformation or compilation and half to execution).  Although
well beyond an acceptable budget for a one-off optimized compilation
procedure, results from exhaustive enumerations are vital for
performance benchmarking of analyses such as ours.  They allow us to
compare our automated procedures, made on the basis of much smaller
searches, with out-of-sample quantities.  They also help build a
library of ``rules of thumb'' and functional and design parameter
characteristics that can be useful for priming future searches whose
tuning variables and input source codes are similar to those of the
fully enumerated experiments.  The fully enumerated GESUMMV problem
(as well as MATMUL and TENSOR) is relatively small from a
performance-tuning perspective, and hence is a prime candidate for our
validation and benchmarking purposes. In [\citet{PBSWBN11}], problems
with up to $10^{53}$ design points are posed, clearly indicating that
practical tuning will require sampling of only a very small portion of
the total design space.

The GESUMMV experiment is summarized as follows. Of the
$2^{3}30^2=7200$ total design points, 199 resulted in a compilation
error or an improper memory access and thus were deemed to violate a
constraint on correctness. The resulting 
245,035 (successful) runtimes were between 0.15 and 0.68 seconds, the
mean and median both being 0.22 seconds. Our focus here is on a
carefully chosen subset of this data, described below, comprising
about 1\% of the full set of runs. The intention is to simulate a
realistic scenario wherein variable selection and sensitivity analysis
techniques can reasonably be expected to add value to an automated
tuning and compilation optimization.

We begin by examining the extent of the cold-cache effect by using
selection techniques.
We then turn to a full analysis of the sensitivity to inputs, leading to
a localization and subsequent optimization.  Next we explore the
extent to which one can learn about, and avoid, constraint violations.
We conclude with an out-of-sample comparison between DTs and GPs.



\begin{figure}

\includegraphics{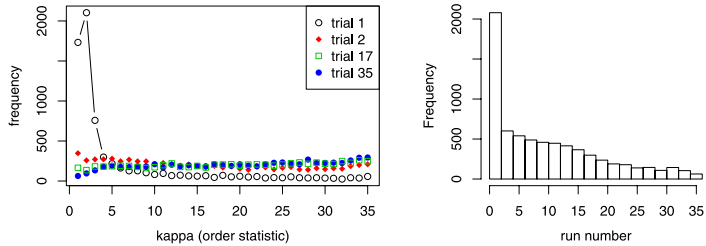}

\caption{(\textit{Left}) Histograms for 4 particular trials with respect to
the order statistics on decreasing runtimes. (\textit{Right}) Frequency
of trial number that yielded the maximum of the 35 runtimes.}
\label{fcoldcache}
\end{figure}

\subsection{Cold-cache effect and variable selection}
\label{secccache}

Figure~\ref{fcoldcache} illustrates the cold-cache effect over the
fully enumerated data.  The left plot shows four histograms counting
the number (out of the 7001 input locations that did not result in a
constraint violation) of times the first, second, seventeenth, and last of
the trials resulted in the $\kappa$th largest runtime of the 35
trials performed.  While the first trial stands out as the slowest,
results for the three other trials indicate that this effect does not
persist for later trials.  That is, the second (17th or 35th) does not
tend to have the second (17th or 35th) largest runtime.  However, the
right histogram in the figure clearly shows that lower trial
numbers tend to yield the maximum execution time more frequently than do
higher ones.

These results make clear the existence of a marginal cold-cache effect;
indeed, a paired $t$-test squarely rejects the null hypothesis that
no marginal effect occurs.  However, our interest lies in
determining whether the effect is influential enough to warrant
inclusion in a model for predicting runtimes.  In particular, the
absolute average distance from the maximum to median runtime (among 35
trials for each of the 7001 input configurations) is about 0.01,
compared with the full difference between the maximum and minimum
execution in the entire data set at 0.53.  Given the effect's low magnitude and
our limited available degrees of freedom, it is not clear
whether estimating the cold-cache effect is worthwhile in
statistical prediction.\looseness=1

The remainder of this subsection and Section~\ref{secsa2} work with
a maxmin space-filling subsampled design of size 500 from the 7001,
and just the first 5 of the 35 replicates (together a 99\% reduction in the
size of the data).
First, we consider the following experiment on a further subset of the
data comprising the first trial and the last (fifth) trial for
every input in the space-filling design (1000 runs in total).  The
five inputs were augmented with a sixth indicator, which is zero for
those from the first trial and one for those from the fifth. If the
cold-cache effect is statistically significant, then this experiment
should reveal so.

\begin{figure}

\includegraphics{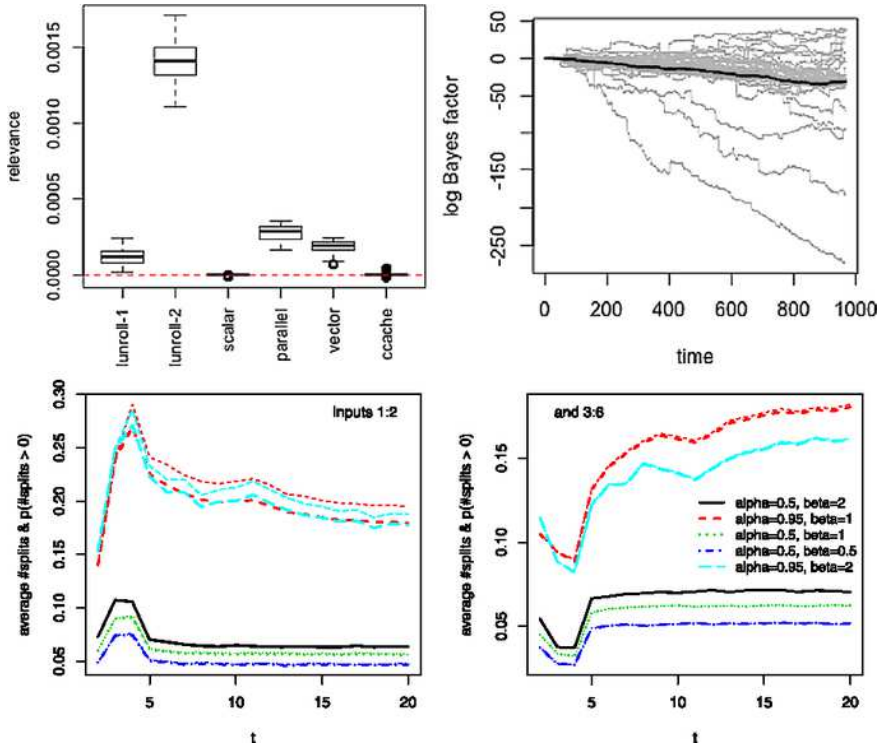}

\caption{(\textit{Top-Left}) Relevance for the five inputs, plus the
cold-cache indicator (sixth input). (\textit{Top-Right}) Sequential
Bayes factors comparing the model with the sixth input to the one
without. (\textit{Bottom}) Prior splitting frequencies (light) and
probabilities of at least one split (dark) for the real-valued
inputs (\textit{left}) and categorical ones (\textit{right}).}
\label{frelevant}
\end{figure}

Figure~\ref{frelevant} summarizes our results, based on a constant
leaf model with 1000 particles and 30 replicates. The top-left panel
shows the posterior relevance samples; the focus, for now, is on the
relevance of the sixth input, which is small. The scale of the
$y$-axis is, however, somewhat deceiving: the posterior probability that
relevance is greater than zero is 0.58, with mean relevance of
$2.6\times 10^{-6}$, indicating that the cold-cache may have a tiny
but possibly significant effect. It is helpful to consult the prior
inclusion probabilities for further guidance here. The bottom-right
figure shows the Boolean predictor's relevance to be approaching 20\%
as the sample size gets large, nearly twice that of our earlier
regression example. The Bayes factor in the the right panel shows a
gradually decreasing trend, signaling that the cold-cache predictor is
not helpful.

Before turning to SA, having decided to ignore the cold-cache effect
based on the above analysis, we observe that input three also shows
low---in fact, negative---relevance.  A similar Bayes factor calculation
(not shown) strongly indicated that it too could be dropped from the
model.  The remaining four inputs have much greater, and entirely
positive, posterior relevance; Bayes factors (also not shown)
reinforce that these predictors are important to obtain a good fit.

\begin{figure}

\includegraphics{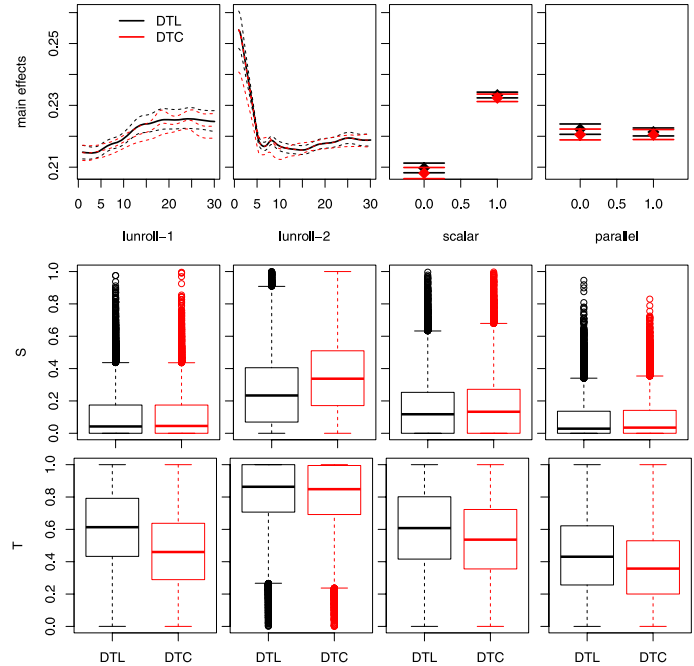}

\caption{Main effects (\textit{first row}) and  $\bm{S}$ (\textit{second row}) and
$\bm{T}$ (\textit{third row}) indices for GESUMMV.}
\label{fgesummvSA}\vspace*{6pt}
\end{figure}

\subsection{Sensitivity analysis}
\label{secsa2}

To further inform an optimization of the automatic code tuning
process, we perform a SA.
Figure~\ref{fgesummvSA} summarizes main effects and $\bm{S}$ and
$\bm{T}$ indices for the four remaining variables.  The full reduced
design (all five trials) was used---25,000 input-output pairs
total, ignoring the cold-cache effect. Results for both constant and
linear leaf models are shown. In contrast to our earlier results
for the Friedman data, the differences between linear and constant
leaves are negligible.  Perhaps this is not surprising since both
treat binary predictors identically.\footnote{It is important to flag the two
remaining binary inputs as
categorical in the \texttt{dynaTree} software when using the linear
leaf model.  This allows splitting on the binary input as usual
but removes such inputs from the within-leaf calculations so that the
resulting Gram matrices are nonsingular.}\vadjust{\goodbreak}

The $\bm{S}$ and $\bm{T}$ indices on the remaining predictors tell a
similar, but richer, story compared with the relevance statistics.
Input two has the largest effect, and input four the smallest, but we
also see that the effect of the inputs, marginally, is small (since
the $S$s are low and the $T$s are high). This result would lead us to
doubt that a rule of thumb for optimizing the codes based on the main
effects alone would bear fruit, namely, that inputs close to $\langle
x_1=5, x_2=12, x_3=0\rangle$ are most promising.  Although this may be
a sensible place to start, intricate interactions among the variables,
as suggested by the $\bm{T}$ indices in particular, mean that a search
for optimal tuning parameters may benefit from a methodical iterative
approach, say, with an expected improvement (EI) criterion
[\citet{jonesschonlauwelch1998}] or another optimization routine.
Before launching headlong in that direction, however, we first
illustrate how a more localized sensitivity analysis may be performed
without revisiting the computations used in the fitting procedure.
The result can either be cached to prime future code optimizations
having similar inputs or to initialize an iterative EI-like search on
a dramatically reduced search space.\looseness=-1

\begin{figure}

\includegraphics{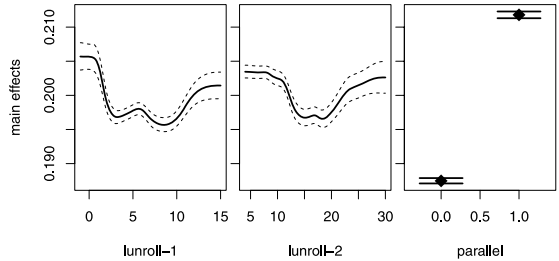}

\caption{Close-up of (constrained) main effects.  $\bm{S}$ and
$\bm{T}$ are similar to Figure~\protect\ref{fgesummvSA}.}
\label{fgesummvSAl}
\end{figure}

Figure~\ref{fgesummvSAl} shows the main effects from a new
sensitivity analysis (using DTC) whose uncertainty distribution
$U'(\bm{x})$ is constrained so that the first input is $\leq 15$, the
second is $\geq 5$, and the third is fixed to zero (representing a tenfold
reduction in the number of possible design points). The relevance
indices indicated importance of the fourth input, so we allowed it to
vary unrestricted in $U'$, suspecting that localizing the first three
inputs might yield a more pronounced effect for the fourth.  Note that
only $U'(\bm{x})$ is restricted, not the actual input-output pairs,
and that the model fitting does not need to be rerun.  In contrast to
the conclusion drawn from Figure~\ref{fgesummvSA},\vadjust{\goodbreak} the localized
analysis strongly indicates that $x_4 = 0$ is required for a locally
optimal solution.  The other two inputs have a smaller marginal
effect, locally.

A finer iterative search may be useful for choosing among the $\geq
2$ local minima in the first two inputs.  Many optimization methods
are viable at this stage.  We prefer to stay within the SMC framework,
allowing thrifty DT updates to pick up where the size 500
space-filling design left off.  Each subsequent design point is chosen
by using a tree-based EI criterion [\citet{taddgrapols2011}] evaluated on
all remaining candidates that meet criteria suggested by our final,
zoomed-in, analysis, namely, all unevaluated locations from the
fully-enumerated set having $\langle x_3, x_4 \rangle = \langle 0, 0
\rangle$ and $\langle x_1, x_2 \rangle \in [2, 12] \times [11,24]$.
Ignoring the irrelevant third input, this results in 98\%
reduction in the search space compared with the original, fully
enumerated design.  After 100 such updates, an evaluation of the
predictive distribution on the full 7001 design led to selecting $x^*
= \langle 4, 22, 0, 0 \rangle$, giving a mean execution time of
$17.9$.

\begin{figure}

\includegraphics{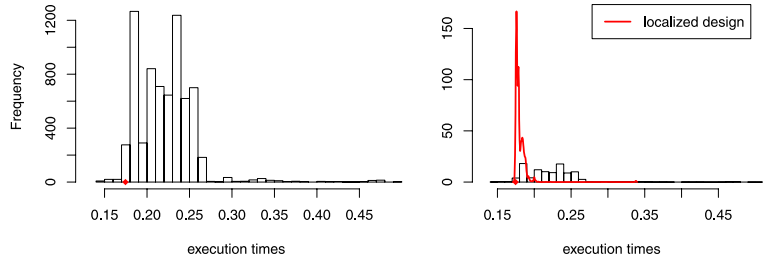}

\caption{Histograms (same on \textit{left} and \textit{right} but with
different $y$-axes) of the median of the 35 runs of each of 7001
non-\texttt{NA} evaluations, shown with the predicted execution time of
$x^*$ found via localized EI (red dot).  The \textit{right} panel
includes a kernel density estimate of the predicted responses at the
localized design, $\langle x_3, x_4 \rangle = \langle 0, 0 \rangle$
and $\langle x_1, x_2 \rangle \in [2, 12] \times [11,24]$.}
\label{fopt}
\end{figure}

Figure~\ref{fopt} shows how this solution is better than 98\% of the
median of the 35 runs from the fully enumerated set.  Both panels show
the same histogram of those times, with a red dot indicating
$\hat{y}(x^*)$.  The right panel augments with the kernel density of
the predicted responses at the reduced/zoomed-in design, indicating the
value of our variable selection and sensitivity pre-analysis.  Even
choosing $x^*$ uniformly at random in this region provides an output
that is better than 88\% of the total options.  The final EI-based
optimization ices the cake.  Since it takes just seconds to perform,
it represents an operation that, when more complete optimization is desired, can
be bolted on at compile time for slight variations of the input source: say, for
differently-sized
matrices.  The ``compiler'' could call up our reduced GESUMMV design
and perform a quick search on the new input matrices.

\subsection{Constraint violation patterns and out-of-sample accuracy}

We return to the original, fully enumerated design to check two
possibilities: (1) whether Argonne engineers unknowingly created an
inefficient timing experiment (i.e., with predictable regions of code
failure); and (2) whether a GP-based analysis would have led to more accurate
predictions, and subsequently a better variable selection, sensitivity
analysis, and optimization, if a vastly greater computational resource had been
available.

For (1), the original 7200 design points, with (two) classification
labels indicating \texttt{NA} values or positive real numbers (times),
were used to fit a DT model with multinomial leaves.  Otherwise the
setup was similar to our earlier examples.  The data has the feature
that if one of the 35 trials was \texttt{NA}, then they all were.  The reason
is that failures were due to the transformed code failing to
compile (precluding the code from running at all) or resulting in a segmentation
fault upon execution. Other failures (e.g., due to hardware failures, soft
faults, or the computed quantities differing more than a certain tolerance from
reference quantities) are possible in practice but were not seen in the present
data.

Sequential updating of the DT classification model revealed
a posterior distribution of the importance indices that was decidedly
null.  The importance probabilities (i.e., of having a positive index)
were $0.003, 0.026, 0.000,\break 0.000$, and $0.000$ for the five inputs,
respectively.  We interpret these results as meaning that the DT model detects
no spatial pattern in the 2.7\% of code failures compared with the
successful runs.  This conclusion was backed up by a simple Bayes
factor calculation where the null model disallowed any partitioning.
These results are reassuring because the input space was designed to
limit the number of correctness violations; if relationships between the
inputs and these violations were known a priori, the
design space would be adjusted accordingly to prevent failures at
compile or runtime.

For (2), we performed a 100-fold Monte Carlo experiment. In each fold,
a DT constant model and a DT linear model were trained on a random
maxmin design of size 100 subsampled from the fully enumerated 7001
locations using the first five replicates.  Besides the smaller
design, the SMC setup is identical to the one described earlier in
this section.  Then, a Bayesian GP with a separable correlation
function and nugget was also fit (using the \texttt{tgp} package) by
using a number of MCMC iterations deemed to give good mixing.  The
resulting computational effort was about ten times greater than for
the DT fit, owing to the $100 \times 100$ matrices that required
repeated inversion.  We originally hoped to do a size 500 design as in
the preceding discussion, but the $2500 \times 2500$ inversions were
computationally infeasible in such nested Monte Carlo repetition.
Finally, a BART model was trained using commensurate MCMC settings.
For validation of the models in each fold a random maxmin testing
design of size 100 was subsampled from the remaining 6901 locations.
RMSEs were obtained by first calculating the squared deviation from
the posterior mean predictors to actual timings of the 35 execution
replicates associated with each testing location. The square root of
the average of the 350 distances was then recorded.\vadjust{\goodbreak}

\begin{figure}
\begin{tabular}{@{}c@{\quad}c@{}}
\begin{minipage}{120pt}
\begin{figure}[H]

\includegraphics{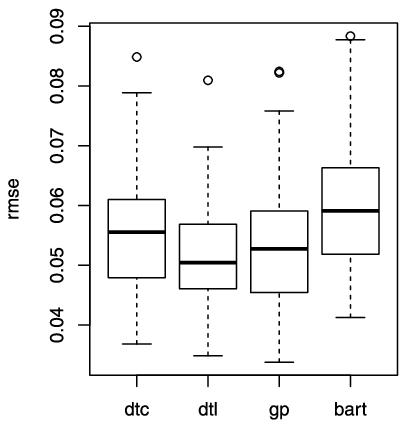}

\end{figure}
\end{minipage}
&
\begin{minipage}{210pt}
\begin{tabular}{@{}lcccc@{}}
\hline
\textbf{RMSE}  &    \textbf{DTC} & \textbf{DTL} & \textbf{GP} & \textbf{BART}\\
\hline
\phantom{0}5\%  & 0.0398 & 0.0373 & 0.0394 & 0.0462 \\
50\% & 0.0525 & 0.0503 & 0.0527 & 0.0591\\
95\% & 0.0675 & 0.0661 & 0.0668 & 0.0750 \\
\hline
\end{tabular}
\vspace*{9.5pt}\end{minipage}
\end{tabular}
\caption{RMSE comparison on GESUMMV Monte Carlo experiment, by boxplot
(\textit{left}) and empirical quantiles (\textit{right}).}\label{frmse}
\end{figure}

Figure~\ref{frmse} summarizes RMSEs by boxplot and numbers: median
and 90\% quantiles. The absolute performance of the DT and GP methods
are strikingly similar. In pairwise comparison, however, DTL is better
than the DTC and GP comparators 85\% and 71\% of the time,
respectively, emerging as a clear winner. Therefore, thrifty
sequential variable selection, sensitivity analysis, and EI-based
optimization notwithstanding, a DT can be at least as good as the
canonical Gaussian process response model for computer experiments.
This performance may be due to a slight nonstationarity or
heteroskedasticity, which cannot be accommodated by the stationary GP.
BART was included as a comparator to further explore this aspect. As
noted by [\citet{taddgrapols2011}], BART will tend to outperform DTs
(and sometimes GPs) when there is nonstationarity or nonsmoothness in
the mean, but not the variance (i.e., under homoskedastic noise). The
opposite is true in the heteroskedastic noise case, and this is what
we observe here. DTC and DTL have lower RMSEs than BART 92\% and 98\%
of the time, respectively. These results suggest our execution-times
data may benefit from methods that can accommodate input-dependent
noise.

\section{Discussion}
\label{secdiscuss}

The advent of fast and cheap computers defined a statistical era in
the late 20th century, especially for Bayesian inference.  For
computer experiments and other spatial data, modestly-sized data sets
and clever algorithms allowed the use of extremely flexible
nonparametric models.  GP models typify the state of the art from that
era, with many successful applications.  In classification problems,
latent variables were key to exploiting computation for modeling
flexibility.  Today, further technological advance is defining a new
era, that of massive data generation and collection where computer and
physical observables are gathered at breakneck pace.

These huge data sets are testing the limits of the popular models and
implementations.  GPs are buckling under the weight of enormous matrix
inverses, and latent variable models suffer from mixing (MCMC)
problems.  Although\vadjust{\goodbreak} exciting inroads have recently been made toward
computationally tractable, approximate GP (regression) inference in
large data settings
[e.g., \citet{haalandqian2012}, \citet{sanghuang2012}], their application
to canonical computer experiments problems such as design and
optimization remains a topic of future study.  In this paper we
suggest that the new method of dynamic regression trees, an update of
classic partition tree techniques, has merit as an efficient
alternative in nonparametric modeling. In particular, we perform many
of the same experiment-analysis functions as do GP and latent variable
models, at a fraction of the computational cost.  By borrowing
relevance statistics from classical trees and sensitivity indices from
GPs, the end product is an exploratory data analysis tool that can
facilitate variable selection, dimension reduction, and visualization.
An open-source implementation is provided in a recent update of the
\texttt{dynaTree} package for \textsf{R}.

Our illustrations included data sets from the recent
literature and a new computer experiment on automatic code generation
that is likely to be a hot application area for statistics and other
disciplines as heterogeneous computing environments become more
commonplace.  Ultimately, the goal is to optimize code for the
architecture ``just in time,'' when it arrives at the computing node.
In order to be realistically achievable, that goal will require
rules of thumb, as facilitated by selection and sensitivity procedures
like those outlined in this paper, and iterative optimization steps
like the EI approach we illustrated.  We note that the input space for
these types of experiments can, in practice, be much larger than the
specific ones we study; indeed, the median size of the problems presented in
[\citet{PBSWBN11}] is more than $10^{15}$ input configurations.  This
makes enumeration prohibitively expensive even for academic purposes,
irregardless of acceptable compilation times.  In those cases,
variable reductions and localizations on the order of those we provide
here will be crucial to enable any study of the search space, let alone a
subsequent optimization.

\begin{appendix}\label{app}

%

\section{Input analysis}

Variable selection is largely equated with setting coefficients to
zero.  Hence, the approaches are predicated on a specific, usually
additive, form for the influence of covariates on response.  In the
analysis of computer experiments, for example,
\citet{cantoniflemmingronchetti2011} and \citet{maitylin2011} use
the nonnegative garrote [NNG, \citet{breiman1995}] and
\citet{huanghorowitzwei2010} apply grouped LASSO.  An advantage of
these approaches is that they can leverage off-the-shelf software for
variable selection.  However, because of the complexity of the modeled
processes and a need for high precision, researchers using statistical
emulation for engineering processes are seldom content with a single
additive regression structure for the entire input space.  Moreover,
the consideration of interaction terms in additive models\vadjust{\goodbreak} can require
huge, overcomplete bases, typically leading to burdensome
computation. As a result, it is more common to rely on GP priors or
other nonparametric regression techniques
[e.g., \citet{BayaBergKennKott2009}, \citet{SansLeeZhouHigd2008}].  However, such
modeling significantly complicates the task of selecting relevant
variables.  Although several approaches have been explored in recent
literature [e.g., \citet{linkletteretal2006}, \citet{bastosohagan2009},
\citet{yishichoi2011}, and references
therein],
their complexity seems to have precluded the release of software for
use by practitioners.

An interesting middle ground is considered by
\citet{reichstorliebondell2009}, who propose an additive model
comprising univariate functions of each predictor and bivariate
functions for all interactive pairs.  Each is given a GP prior, and
there is a catch-all (higher interactions) remainder term.  This
extends previous work wherein $B$-splines were proposed for a similar
task [e.g., \citet{gu2002}].  Stochastic-search variable selection
[SSVS, \citet{georgemccu1993}] is used for selecting main effects
and interactions.  Although perhaps more straightforward than performing
SSVS directly on the lengthscale parameters of a GP
[\citet{linkletteretal2006}], this approach has the added
computational complexity of inverting many $(\mathcal{O}(m^2)) n \times n$
covariance matrices.

Instead of a dedicated variable selection procedure, engineering
simulators typically employ some form of input sensitivity analysis.
Classically, as in examples from \citet{SaltChanScot2000}, \citet{SaltEtAl2008},
running the computer code to obtain a response is presumed to be
cheap.  When it is expensive, one must emulate the code
with an estimated probability model [see \citet{santwillnotz2003} for an
overview]. In turn, researchers have proposed a
variety of schemes for extension of classic sensitivity analysis to
account for response surface uncertainty.  GPs, because of their role as
the canonical choice for modeling computer experiments, are
combined with sensitivity analysis in applications
[e.g., \citet{ziehntomlin2009,marelletal2009}].  However, the
associated methodology is usually based on restrictive stationarity
and homoskedasticity assumptions needed to derive either empirical
Bayes [\citet{oakleyohagan2004}] or fully Bayesian
[\citet{MorrKottTaddFurfGana2008,farahkottas2011}] estimates of
sensitivity indices.  Notable exceptions are presented by
\citet{storlieetal2009} and \citet{TaddLeeGrayGrif2009}.  In the
former, approximate bootstrap confidence intervals are derived for
sensitivity indices based on a nonparametrically modeled response
surface.  In the latter, variability integration embedded within MCMC
simulation yields samples from sensitivity indices' full posterior
distribution; a similar idea forms the basis for our framework in
Section~\ref{secSA}.

Partition trees [e.g., CART: \citet{brei1984}] provide a basis for
regression that has both a simplicity amenable to selecting variables
and the flexibility required for modeling computer experiments.
Furthermore, partition trees overcome some well-known drawbacks of the
more commonly applied GP computer emulators: expensive
$\mathcal{O}(n^3)$ matrix inversion, involving special consideration\vadjust{\goodbreak}
for categorical predictors and responses and allowing for the possibility
of nonstationarity in the response or heteroskedastic errors.
\citet{taddgrapols2011} provide extensive background on general
tree-based regression and argue for its wider adoption in engineering
applications.  In the context of this paper's goals, trees present a
unique, nonadditive foundation for determining variable relevance.  In
their simplest form, with constant mean response at the tree
leaves, variable selection is automatic: if a variable is never used
to define a tree partition, it has been effectively removed from the
regression.  Indeed, this idea motivated some of the earliest work on
the use of trees, as presented by \citet{MorgSonq1963}, for automatic interaction
detection.  In a more nuanced approach, \citet{brei1984} introduce
indices of variable importance that measure squared error reduction
due to tree-splits defined on each covariate.
\citeauthor{hastietibshfried2009} [HTF; (\citeyear{hastietibshfried2009}), Chapter~10] promote these indices
for sensitivity analysis and describe how the approach can be extended
for their boosted trees.

However, these techniques are purely algorithmic and lack a full
probability model, hence, their use is especially problematic in
analysis of computer experiments, where uncertainty quantification is
often a primary objective.  Moreover, the HTF importance indices are
only point estimates of the underlying sensitivity metrics, thus, they
preclude basing the variable selection criteria on posterior evidence
and make it difficult to properly deduce and interpret just how each
variable is contributing.  Researchers have attempted to overcome some
of these shortcomings through the use of Bayesian inference, most
recently in schemes that augment the tree model to allow for better
control or flexibility. \citet{ChipGeorMcCu2010} describe a Bayesian
additive regression tree (BART) model, and their \texttt{BayesTree}
software includes a direct analogue of the HTF importance indices; and
the method of \citet{TaddLeeGrayGrif2009} is implemented in \texttt{tgp}
[see \citet{gramacytaddy2010}, Section~3].


\section{Variance integral for linear leaves}
\label{seclinint}

Here, we derive the variance integrals from (\ref{eqredvar}) for a
model with linear leaves.  Dropping the node subscript ($\eta$,
$\eta_\ell$, or $\eta_r$), we have
\begin{eqnarray}\label{eqs2lin}
\int_A \sigma^2(\bm{x}) \,d\bm{x} &=& \int
_A \frac{s^2 - \mathcal{R}}{n-p-1} \biggl(1 + \frac{1}{n} +
\bm{x}^\top \mathcal{G}^{-1} \bm{x} \biggr) \,d\bm{x}
\nonumber
\\[-8pt]
\\[-8pt]
\nonumber
&= &\frac{s^2 - \mathcal{R}}{n-p-1} \biggl(|A| \biggl(1 + \frac{1}{n} \biggr) + \int
_A \bm{x}^\top \mathcal{G}^{-1} \bm{x}
\,d\bm{x} \biggr),
\nonumber
\end{eqnarray}
where $s^2$ is the sum of squares, $\mathcal{R}$ is the regression sum
of squares, $n\equiv|\eta|$ is the number of $(\bm{x},y)$ pairs,
$\mathcal{G}$ is the Gram matrix, and $|A|$ is the area of the
rectangle.  The remaining integral is just a sum of polynomials: with
the intervals outlining\vadjust{\goodbreak} the rectangle given by $(a_1,b_1),\ldots,(a_p,
b_p)$ and $(g_{ij})$ the components of $\mathcal{G}^{-1}$,
\begin{eqnarray}\label{eqint}
\int_A \bm{x}^\top \mathcal{G}^{-1}
\bm{x} \,d\bm{x} &=& \int_{a_1}^{b_1} \cdots \int
_{a_p}^{b_p} \sum_{i=1}^p
\sum_{j=1}^p x_i
x_j g_{ij} \,d x_i
\nonumber\\[-2pt]
&=& \sum_{i=1}^p \frac{g_{ii}}{3}
\bigl(b_i^3 - a_i^3\bigr) \prod
_{k\ne i} (b_k - a_k)
\nonumber
\\[-9pt]
\\[-9pt]
\nonumber
&&{}+ 2 \sum
_{i=1}^p \sum_{j > i}
\frac{g_{ij}}{4}\bigl(b_i^2 - a_i^2
\bigr) \bigl(b_j^2 - a_j^2\bigr)
\prod_{k
\ne i,j} (b_k - a_k)
\\[-2pt]
&=& |A| \Biggl( \sum
_{i=1}^p \frac{g_{ii}(b_i^3 - a_i^3)}{3  (b_i - a_i)} + \sum
_{i=1}^p \sum_{j > i}
\frac{g_{ij} (b_i+ a_i)(b_j+a_j)}{2} \Biggr).
\nonumber
\end{eqnarray}
\end{appendix}

\section*{Acknowledgments}\label{secack}
We are grateful to Prasanna Balaprakash for providing the data from
[\citet{Balaprakash20112136}]. Many
thanks to the Editor, Associate Editor, and two referees for their
valuable comments, which led to many improvements.


\printaddresses


\begin{thebibliography}{49}

\bibitem[\protect\citeauthoryear{Asuncion and
  Newman}{2007}]{Asuncion+Newman2007}
\begin{bmisc}[author]
\bauthor{\bsnm{Asuncion},~\bfnm{A.}\binits{A.}} \AND
  \bauthor{\bsnm{Newman},~\bfnm{D.~J.}\binits{D.~J.}}
(\byear{2007}).
\bhowpublished{{UCI} machine learning repository. Available
  at \url{http://www.ics.uci.edu/\textasciitilde mlearn/MLRepository.html}.}
\bptok{imsref}%
\end{bmisc}
\endbibitem

\bibitem[\protect\citeauthoryear{Balaprakash, Wild and
  Hovland}{2011}]{Balaprakash20112136}
\begin{barticle}[author]
\bauthor{\bsnm{Balaprakash},~\bfnm{Prasanna}\binits{P.}},
  \bauthor{\bsnm{Wild},~\bfnm{Stefan~M.}\binits{S.~M.}} \AND
  \bauthor{\bsnm{Hovland},~\bfnm{Paul~D.}\binits{P.~D.}}
(\byear{2011}).
\btitle{Can search algorithms save large-scale automatic performance tuning?}
\bjournal{Procedia Computer Science}
\bvolume{4}
\bpages{2136--2145}.
\bptok{imsref}%
\end{barticle}
\endbibitem

\bibitem[\protect\citeauthoryear{Balaprakash, Wild and Norris}{2012}]{PBSWBN11}
\begin{barticle}[author]
\bauthor{\bsnm{Balaprakash},~\bfnm{Prasanna}\binits{P.}},
  \bauthor{\bsnm{Wild},~\bfnm{Stefan~M.}\binits{S.~M.}} \AND
  \bauthor{\bsnm{Norris},~\bfnm{Boyana}\binits{B.}}
(\byear{2012}).
\btitle{SPAPT: Search problems in automatic performance tuning}.
\bjournal{Procedia Computer Science}
\bvolume{9}
\bpages{1959--1968}.
\bptok{imsref}%
\end{barticle}
\endbibitem

\bibitem[\protect\citeauthoryear{Bastos and O'Hagan}{2009}]{bastosohagan2009}
\begin{barticle}[mr]
\bauthor{\bsnm{Bastos},~\bfnm{Leonardo~S.}\binits{L.~S.}} \AND
  \bauthor{\bsnm{O'Hagan},~\bfnm{Anthony}\binits{A.}}
(\byear{2009}).
\btitle{Diagnostics for {G}aussian process emulators}.
\bjournal{Technometrics}
\bvolume{51}
\bpages{425--438}.
\bid{doi={10.1198/TECH.2009.08019}, issn={0040-1706}, mr={2756478}}
\bptok{imsref}%
\end{barticle}
\endbibitem

\bibitem[\protect\citeauthoryear{Bayarri et~al.}{2009}]{BayaBergKennKott2009}
\begin{barticle}[mr]
\bauthor{\bsnm{Bayarri},~\bfnm{M.~J.}\binits{M.~J.}},
  \bauthor{\bsnm{Berger},~\bfnm{James~O.}\binits{J.~O.}},
  \bauthor{\bsnm{Kennedy},~\bfnm{Marc~C.}\binits{M.~C.}},
  \bauthor{\bsnm{Kottas},~\bfnm{Athanasios}\binits{A.}},
  \bauthor{\bsnm{Paulo},~\bfnm{Rui}\binits{R.}},
  \bauthor{\bsnm{Sacks},~\bfnm{Jerry}\binits{J.}},
  \bauthor{\bsnm{Cafeo},~\bfnm{John~A.}\binits{J.~A.}},
  \bauthor{\bsnm{Lin},~\bfnm{Chin-Hsu}\binits{C.-H.}} \AND
  \bauthor{\bsnm{Tu},~\bfnm{Jian}\binits{J.}}
(\byear{2009}).
\btitle{Predicting vehicle crashworthiness: Validation of computer models for
  functional and hierarchical data}.
\bjournal{J. Amer. Statist. Assoc.}
\bvolume{104}
\bpages{929--943}.
\bid{doi={10.1198/jasa.2009.ap06623}, issn={0162-1459}, mr={2750226}}
\bptok{imsref}%
\end{barticle}
\endbibitem

\bibitem[\protect\citeauthoryear{Blackford et~al.}{2002}]{UBLAS}
\begin{barticle}[mr]
\bauthor{\bsnm{Blackford},~\bfnm{L.~Susan}\binits{L.~S.}},
  \bauthor{\bsnm{Demmel},~\bfnm{J.}\binits{J.}},
  \bauthor{\bsnm{Dongarra},~\bfnm{J.}\binits{J.}},
  \bauthor{\bsnm{Duff},~\bfnm{I.}\binits{I.}},
  \bauthor{\bsnm{Hammarling},~\bfnm{S.}\binits{S.}},
  \bauthor{\bsnm{Henry},~\bfnm{G.}\binits{G.}},
  \bauthor{\bsnm{Heroux},~\bfnm{M.}\binits{M.}},
  \bauthor{\bsnm{Kaufman},~\bfnm{L.}\binits{L.}},
  \bauthor{\bsnm{Lumsdaine},~\bfnm{A.}\binits{A.}},
  \bauthor{\bsnm{Petitet},~\bfnm{A.}\binits{A.}},
  \bauthor{\bsnm{Pozo},~\bfnm{R.}\binits{R.}},
  \bauthor{\bsnm{Remington},~\bfnm{K.}\binits{K.}} \AND
  \bauthor{\bsnm{Whaley},~\bfnm{R.~C.}\binits{R.~C.}}
(\byear{2002}).
\btitle{An updated set of basic linear algebra subprograms ({BLAS})}.
\bjournal{ACM Trans. Math. Software}
\bvolume{28}
\bpages{135--151}.
\bid{doi={10.1145/567806.567807}, issn={0098-3500}, mr={1928065}}
\bptok{imsref}%
\end{barticle}
\endbibitem

\bibitem[\protect\citeauthoryear{Breiman}{1995}]{breiman1995}
\begin{barticle}[mr]
\bauthor{\bsnm{Breiman},~\bfnm{Leo}\binits{L.}}
(\byear{1995}).
\btitle{Better subset regression using the nonnegative garrote}.
\bjournal{Technometrics}
\bvolume{37}
\bpages{373--384}.
\bid{doi={10.2307/1269730}, issn={0040-1706}, mr={1365720}}
\bptok{imsref}%
\end{barticle}
\endbibitem

\bibitem[\protect\citeauthoryear{Breiman et~al.}{1984}]{brei1984}
\begin{bbook}[author]
\bauthor{\bsnm{Breiman},~\bfnm{L.}\binits{L.}},
  \bauthor{\bsnm{Friedman},~\bfnm{J.~H.}\binits{J.~H.}},
  \bauthor{\bsnm{Olshen},~\bfnm{R.}\binits{R.}} \AND
  \bauthor{\bsnm{Stone},~\bfnm{C.}\binits{C.}}
(\byear{1984}).
\btitle{Classification and Regression Trees}.
\bpublisher{Wadsworth}, \blocation{Belmont, CA}.
\bptok{imsref}%
\end{bbook}
\endbibitem

\bibitem[\protect\citeauthoryear{Cantoni, Flemming and
  Ronchetti}{2011}]{cantoniflemmingronchetti2011}
\begin{barticle}[mr]
\bauthor{\bsnm{Cantoni},~\bfnm{Eva}\binits{E.}},
  \bauthor{\bsnm{Flemming},~\bfnm{Joanna~Mills}\binits{J.~M.}} \AND
  \bauthor{\bsnm{Ronchetti},~\bfnm{Elvezio}\binits{E.}}
(\byear{2011}).
\btitle{Variable selection in additive models by non-negative garrote}.
\bjournal{Stat. Model.}
\bvolume{11}
\bpages{237--252}.
\bid{doi={10.1177/1471082X1001100304}, issn={1471-082X}, mr={2857594}}
\bptok{imsref}%
\end{barticle}
\endbibitem

\bibitem[\protect\citeauthoryear{Carvalho et~al.}{2010}]{CarvJohaLopePols2009}
\begin{barticle}[mr]
\bauthor{\bsnm{Carvalho},~\bfnm{Carlos~M.}\binits{C.~M.}},
  \bauthor{\bsnm{Johannes},~\bfnm{Michael~S.}\binits{M.~S.}},
  \bauthor{\bsnm{Lopes},~\bfnm{Hedibert~F.}\binits{H.~F.}} \AND
  \bauthor{\bsnm{Polson},~\bfnm{Nicholas~G.}\binits{N.~G.}}
(\byear{2010}).
\btitle{Particle learning and smoothing}.
\bjournal{Statist. Sci.}
\bvolume{25}
\bpages{88--106}.
\bid{doi={10.1214/10-STS325}, issn={0883-4237}, mr={2741816}}
\bptok{imsref}%
\end{barticle}
\endbibitem

\bibitem[\protect\citeauthoryear{Chipman, George and
  Mc{C}ulloch}{1998}]{chipgeormccu1998}
\begin{barticle}[author]
\bauthor{\bsnm{Chipman},~\bfnm{H.~A.}\binits{H.~A.}},
  \bauthor{\bsnm{George},~\bfnm{E.~I.}\binits{E.~I.}} \AND
  \bauthor{\bsnm{Mc{C}ulloch},~\bfnm{R.~E.}\binits{R.~E.}}
(\byear{1998}).
\btitle{{B}ayesian {CART} model search (with discussion)}.
\bjournal{J. Amer. Statist. Assoc.}
\bvolume{93}
\bpages{935--960}.
\bptok{imsref}%
\end{barticle}\vadjust{\goodbreak}
\endbibitem

\bibitem[\protect\citeauthoryear{Chipman, George and
  Mc{C}ulloch}{2002}]{chipgeormccu2002}
\begin{barticle}[author]
\bauthor{\bsnm{Chipman},~\bfnm{H.~A.}\binits{H.~A.}},
  \bauthor{\bsnm{George},~\bfnm{E.~I.}\binits{E.~I.}} \AND
  \bauthor{\bsnm{Mc{C}ulloch},~\bfnm{R.~E.}\binits{R.~E.}}
(\byear{2002}).
\btitle{Bayesian treed models}.
\bjournal{Machine Learning}
\bvolume{48}
\bpages{303--324}.
\bptok{imsref}%
\end{barticle}
\endbibitem

\bibitem[\protect\citeauthoryear{Chipman, George and
  McCulloch}{2010}]{ChipGeorMcCu2010}
\begin{barticle}[mr]
\bauthor{\bsnm{Chipman},~\bfnm{Hugh~A.}\binits{H.~A.}},
  \bauthor{\bsnm{George},~\bfnm{Edward~I.}\binits{E.~I.}} \AND
  \bauthor{\bsnm{McCulloch},~\bfnm{Robert~E.}\binits{R.~E.}}
(\byear{2010}).
\btitle{B{ART}: {B}ayesian additive regression trees}.
\bjournal{Ann. Appl. Stat.}
\bvolume{4}
\bpages{266--298}.
\bid{doi={10.1214/09-AOAS285}, issn={1932-6157}, mr={2758172}}
\bptok{imsref}%
\end{barticle}
\endbibitem

\bibitem[\protect\citeauthoryear{Farah and Kottas}{2011}]{farahkottas2011}
\begin{bmisc}[author]
\bauthor{\bsnm{Farah},~\bfnm{M.}\binits{M.}} \AND
  \bauthor{\bsnm{Kottas},~\bfnm{A.}\binits{A.}}
(\byear{2011}).
\bhowpublished{Bayesian inference for sensitivity analysis of computer
  simulators, with an application to radiative transfer models. Technical
  Report UCSC-SOE-10-15, Univ. California, Santa Cruz.}
\bptok{imsref}%
\end{bmisc}
\endbibitem

\bibitem[\protect\citeauthoryear{Friedman}{1991}]{fried1991}
\begin{barticle}[mr]
\bauthor{\bsnm{Friedman},~\bfnm{Jerome~H.}\binits{J.~H.}}
(\byear{1991}).
\btitle{Multivariate adaptive regression splines}.
\bjournal{Ann. Statist.}
\bvolume{19}
\bpages{1--141}.
\bid{doi={10.1214/aos/1176347963}, issn={0090-5364}, mr={1091842}}
\bptnote{check related}%
\bptok{imsref}%
\end{barticle}
\endbibitem

\bibitem[\protect\citeauthoryear{Friedman, Hastie and
  Tibshirani}{2010}]{friedhasttibsh2009}
\begin{barticle}[author]
\bauthor{\bsnm{Friedman},~\bfnm{Jerome~H.}\binits{J.~H.}},
  \bauthor{\bsnm{Hastie},~\bfnm{Trevor}\binits{T.}} \AND
  \bauthor{\bsnm{Tibshirani},~\bfnm{Rob}\binits{R.}}
(\byear{2010}).
\btitle{Regularization paths for generalized linear models via coordinate
  descent}.
\bjournal{Journal of Statistical Software}
\bvolume{33}
\bpages{1--22}.
\bptok{imsref}%
\end{barticle}
\endbibitem

\bibitem[\protect\citeauthoryear{George and Mc{C}ulloch}{1993}]{georgemccu1993}
\begin{barticle}[author]
\bauthor{\bsnm{George},~\bfnm{E.~I.}\binits{E.~I.}} \AND
  \bauthor{\bsnm{Mc{C}ulloch},~\bfnm{R.~E.}\binits{R.~E.}}
(\byear{1993}).
\btitle{Variable selection via {G}ibbs sampling}.
\bjournal{J.~Amer. Statist. Assoc.}
\bvolume{88}
\bpages{881--889}.
\bptok{imsref}%
\end{barticle}
\endbibitem

\bibitem[\protect\citeauthoryear{Gramacy and Polson}{2011}]{gramacypolson2011}
\begin{barticle}[mr]
\bauthor{\bsnm{Gramacy},~\bfnm{Robert~B.}\binits{R.~B.}} \AND
  \bauthor{\bsnm{Polson},~\bfnm{Nicholas~G.}\binits{N.~G.}}
(\byear{2011}).
\btitle{Particle learning of {G}aussian process models for sequential design
  and optimization}.
\bjournal{J. Comput. Graph. Statist.}
\bvolume{20}
\bpages{102--118}.
\bid{doi={10.1198/jcgs.2010.09171}, issn={1061-8600}, mr={2816540}}
\bptok{imsref}%
\end{barticle}
\endbibitem

\bibitem[\protect\citeauthoryear{Gramacy and Polson}{2012}]{grapols2012}
\begin{barticle}[author]
\bauthor{\bsnm{Gramacy},~\bfnm{R.~B.}\binits{R.~B.}} \AND
  \bauthor{\bsnm{Polson},~\bfnm{N.~G.}\binits{N.~G.}}
(\byear{2012}).
\btitle{Simulation-based regularized logistic regression}.
\bjournal{Bayesian Anal.}
\bvolume{7}
\bpages{1--24}.
\bptok{imsref}%
\end{barticle}
\endbibitem

\bibitem[\protect\citeauthoryear{Gramacy and Taddy}{2010}]{gramacytaddy2010}
\begin{barticle}[author]
\bauthor{\bsnm{Gramacy},~\bfnm{Robert~B.}\binits{R.~B.}} \AND
  \bauthor{\bsnm{Taddy},~\bfnm{Matthew~Alan}\binits{M.~A.}}
(\byear{2010}).
\btitle{Categorical inputs, sensitivity analysis, optimization and importance
  tempering with \texttt{tgp} version 2, an \textsf{R} package for treed
  Gaussian process models}.
\bjournal{Journal of Statistical Software}
\bvolume{33}
\bpages{1--48}.
\bptok{imsref}%
\end{barticle}
\endbibitem

\bibitem[\protect\citeauthoryear{Gramacy and Taddy}{2011}]{dynaTree}
\begin{bmisc}[author]
\bauthor{\bsnm{Gramacy},~\bfnm{Robert~B.}\binits{R.~B.}} \AND
  \bauthor{\bsnm{Taddy},~\bfnm{Matt~A.}\binits{M.~A.}}
(\byear{2011}).
\bhowpublished{\texttt{dynaTree}: {D}ynamic trees for learning and design. R
  package version 2.0}.
\bptok{imsref}%
\end{bmisc}
\endbibitem

\bibitem[\protect\citeauthoryear{Gu}{2002}]{gu2002}
\begin{bbook}[mr]
\bauthor{\bsnm{Gu},~\bfnm{Chong}\binits{C.}}
(\byear{2002}).
\btitle{Smoothing Spline {ANOVA} Models}.
\bpublisher{Springer}, \blocation{New York}.
\bid{mr={1876599}}
\bptok{imsref}%
\end{bbook}
\endbibitem

\bibitem[\protect\citeauthoryear{Haaland and Qian}{2011}]{haalandqian2012}
\begin{barticle}[author]
\bauthor{\bsnm{Haaland},~\bfnm{B.}\binits{B.}} \AND
  \bauthor{\bsnm{Qian},~\bfnm{P.~Z.~G.}\binits{P.~Z.~G.}}
(\byear{2011}).
\btitle{Accurate emulators for large-scale computer experiments}.
\bjournal{Ann. Statist.}
\bvolume{39}
\bpages{2974--3002}.
\bptok{imsref}%
\end{barticle}
\endbibitem

\bibitem[\protect\citeauthoryear{Hartono, Norris and Sadayappan}{2009}]{Orio09}
\begin{binproceedings}[author]
\bauthor{\bsnm{Hartono},~\bfnm{A.}\binits{A.}},
  \bauthor{\bsnm{Norris},~\bfnm{B.}\binits{B.}} \AND
  \bauthor{\bsnm{Sadayappan},~\bfnm{P.}\binits{P.}}
(\byear{2009}).
\btitle{Annotation-based empirical performance tuning using orio}.
In \bbooktitle{Proceedings of the IEEE International Symposium on Parallel
  Distributed Processing, 2009 (IPDPS 2009)}
\bpages{1--11}.
\bpublisher{IEEE}, \blocation{New York}.
\bptok{imsref}%
\end{binproceedings}
\endbibitem

\bibitem[\protect\citeauthoryear{Hastie, Tibshirani and
  Friedman}{2009}]{hastietibshfried2009}
\begin{bbook}[mr]
\bauthor{\bsnm{Hastie},~\bfnm{Trevor}\binits{T.}},
  \bauthor{\bsnm{Tibshirani},~\bfnm{Robert}\binits{R.}} \AND
  \bauthor{\bsnm{Friedman},~\bfnm{Jerome}\binits{J.}}
(\byear{2009}).
\btitle{The Elements of Statistical Learning: Data Mining, Inference, and
  Prediction},
\bedition{2nd} ed.
\bpublisher{Springer}, \blocation{New York}.
\bid{doi={10.1007/978-0-387-84858-7}, mr={2722294}}
\bptok{imsref}%
\end{bbook}
\endbibitem

\bibitem[\protect\citeauthoryear{Huang, Horowitz and
  Wei}{2010}]{huanghorowitzwei2010}
\begin{barticle}[mr]
\bauthor{\bsnm{Huang},~\bfnm{Jian}\binits{J.}},
  \bauthor{\bsnm{Horowitz},~\bfnm{Joel~L.}\binits{J.~L.}} \AND
  \bauthor{\bsnm{Wei},~\bfnm{Fengrong}\binits{F.}}
(\byear{2010}).
\btitle{Variable selection in nonparametric additive models}.
\bjournal{Ann. Statist.}
\bvolume{38}
\bpages{2282--2313}.
\bid{doi={10.1214/09-AOS781}, issn={0090-5364}, mr={2676890}}
\bptok{imsref}%
\end{barticle}
\endbibitem

\bibitem[\protect\citeauthoryear{Jones, Schonlau and
  Welch}{1998}]{jonesschonlauwelch1998}
\begin{barticle}[mr]
\bauthor{\bsnm{Jones},~\bfnm{Donald~R.}\binits{D.~R.}},
  \bauthor{\bsnm{Schonlau},~\bfnm{Matthias}\binits{M.}} \AND
  \bauthor{\bsnm{Welch},~\bfnm{William~J.}\binits{W.~J.}}
(\byear{1998}).
\btitle{Efficient global optimization of expensive black-box functions}.
\bjournal{J. Global Optim.}
\bvolume{13}
\bpages{455--492}.
\bid{doi={10.1023/A:1008306431147}, issn={0925-5001}, mr={1673460}}
\bptok{imsref}%
\end{barticle}
\endbibitem

\bibitem[\protect\citeauthoryear{Krishnapuram et~al.}{2005}]{krishetal2005}
\begin{barticle}[author]
\bauthor{\bsnm{Krishnapuram},~\bfnm{B.}\binits{B.}},
  \bauthor{\bsnm{Carin},~\bfnm{L.}\binits{L.}},
  \bauthor{\bsnm{Figueiredo},~\bfnm{M.}\binits{M.}} \AND
  \bauthor{\bsnm{Hartemink},~\bfnm{A.}\binits{A.}}
(\byear{2005}).
\btitle{Sparse multinomial logistic regression: Fast algorithms and
  generalization bounds}.
\bjournal{IEEE Transactions on Pattern Analysis and Machine Intelligence}
\bvolume{27}
\bpages{957--969}.
\bptok{imsref}%
\end{barticle}
\endbibitem

\bibitem[\protect\citeauthoryear{Lee et~al.}{2008}]{SansLeeZhouHigd2008}
\begin{barticle}[mr]
\bauthor{\bsnm{Lee},~\bfnm{Herbert K.~H.}\binits{H.~K.~H.}},
  \bauthor{\bsnm{Sans{\'o}},~\bfnm{Bruno}\binits{B.}},
  \bauthor{\bsnm{Zhou},~\bfnm{Weining}\binits{W.}} \AND
  \bauthor{\bsnm{Higdon},~\bfnm{David~M.}\binits{D.~M.}}
(\byear{2008}).
\btitle{Inference for a proton accelerator using convolution models}.
\bjournal{J. Amer. Statist. Assoc.}
\bvolume{103}
\bpages{604--613}.
\bid{doi={10.1198/016214507000000833}, issn={0162-1459}, mr={2523997}}
\bptok{imsref}%
\end{barticle}
\endbibitem

\bibitem[\protect\citeauthoryear{Linkletter et~al.}{2006}]{linkletteretal2006}
\begin{barticle}[mr]
\bauthor{\bsnm{Linkletter},~\bfnm{Crystal}\binits{C.}},
  \bauthor{\bsnm{Bingham},~\bfnm{Derek}\binits{D.}},
  \bauthor{\bsnm{Hengartner},~\bfnm{Nicholas}\binits{N.}},
  \bauthor{\bsnm{Higdon},~\bfnm{David}\binits{D.}} \AND
  \bauthor{\bsnm{Ye},~\bfnm{Kenny~Q.}\binits{K.~Q.}}
(\byear{2006}).
\btitle{Variable selection for {G}aussian process models in computer
  experiments}.
\bjournal{Technometrics}
\bvolume{48}
\bpages{478--490}.
\bid{doi={10.1198/004017006000000228}, issn={0040-1706}, mr={2328617}}
\bptok{imsref}%
\end{barticle}
\endbibitem

\bibitem[\protect\citeauthoryear{Maity and Lin}{2011}]{maitylin2011}
\begin{barticle}[mr]
\bauthor{\bsnm{Maity},~\bfnm{Arnab}\binits{A.}} \AND
  \bauthor{\bsnm{Lin},~\bfnm{Xihong}\binits{X.}}
(\byear{2011}).
\btitle{Powerful tests for detecting a gene effect in the presence of possible
  gene-gene interactions using garrote kernel machines}.
\bjournal{Biometrics}
\bvolume{67}
\bpages{1271--1284}.
\bid{doi={10.1111/j.1541-0420.2011.01598.x}, issn={0006-341X}, mr={2872377}}
\bptok{imsref}%
\end{barticle}
\endbibitem

\bibitem[\protect\citeauthoryear{Marrel et~al.}{2009}]{marelletal2009}
\begin{barticle}[author]
\bauthor{\bsnm{Marrel},~\bfnm{A.}\binits{A.}},
  \bauthor{\bsnm{Iooss},~\bfnm{B.}\binits{B.}},
  \bauthor{\bsnm{Laurent},~\bfnm{B.}\binits{B.}} \AND
  \bauthor{\bsnm{Roustant},~\bfnm{O.}\binits{O.}}
(\byear{2009}).
\btitle{Calculations of Sobol indices for the Gaussian process metamodel}.
\bjournal{Reliability Engineering and System Safety}
\bvolume{94}
\bpages{742--751}.
\bptok{imsref}%
\end{barticle}
\endbibitem

\bibitem[\protect\citeauthoryear{Morgan and Sonquist}{1963}]{MorgSonq1963}
\begin{barticle}[author]
\bauthor{\bsnm{Morgan},~\bfnm{James~N.}\binits{J.~N.}} \AND
  \bauthor{\bsnm{Sonquist},~\bfnm{John~A.}\binits{J.~A.}}
(\byear{1963}).
\btitle{Problems in the analysis of survey data, and a proposal}.
\bjournal{J. Amer. Statist. Assoc.}
\bvolume{58}
\bpages{415--434}.
\bptok{imsref}%
\end{barticle}
\endbibitem

\bibitem[\protect\citeauthoryear{Morris
  et~al.}{2008}]{MorrKottTaddFurfGana2008}
\begin{barticle}[author]
\bauthor{\bsnm{Morris},~\bfnm{R.~D.}\binits{R.~D.}},
  \bauthor{\bsnm{Kottas},~\bfnm{A.}\binits{A.}},
  \bauthor{\bsnm{Taddy},~\bfnm{M.}\binits{M.}},
  \bauthor{\bsnm{Furfaro},~\bfnm{R.}\binits{R.}} \AND
  \bauthor{\bsnm{Ganapol},~\bfnm{B.}\binits{B.}}
(\byear{2008}).
\btitle{A~statistical framework for the sensitivity analysis of radiative
  transfer models}.
\bjournal{IEEE Transactions on Geoscience and Remote Sensing}
\bvolume{12}
\bpages{4062--4074}.
\bptok{imsref}%
\end{barticle}
\endbibitem

\bibitem[\protect\citeauthoryear{Oakley and O'Hagan}{2004}]{oakleyohagan2004}
\begin{barticle}[mr]
\bauthor{\bsnm{Oakley},~\bfnm{Jeremy~E.}\binits{J.~E.}} \AND
  \bauthor{\bsnm{O'Hagan},~\bfnm{Anthony}\binits{A.}}
(\byear{2004}).
\btitle{Probabilistic sensitivity analysis of complex models: A~{B}ayesian
  approach}.
\bjournal{J. R. Stat. Soc. Ser. B Stat. Methodol.}
\bvolume{66}
\bpages{751--769}.
\bid{doi={10.1111/j.1467-9868.2004.05304.x}, issn={1369-7412}, mr={2088780}}
\bptok{imsref}%
\end{barticle}
\endbibitem

\bibitem[\protect\citeauthoryear{Patterson and Hennessy}{2007}]{DPJLbook}
\begin{bbook}[author]
\bauthor{\bsnm{Patterson},~\bfnm{David~A.}\binits{D.~A.}} \AND
  \bauthor{\bsnm{Hennessy},~\bfnm{John~L.}\binits{J.~L.}}
(\byear{2007}).
\btitle{Computer Organization and Design---the Hardware / Software Interface},
\bedition{3rd} ed.
\bpublisher{Morgan Kaufmann}, \blocation{Boston}.
\bptok{imsref}%
\end{bbook}
\endbibitem

\bibitem[\protect\citeauthoryear{Reich, Storlie and
  Bondell}{2009}]{reichstorliebondell2009}
\begin{barticle}[mr]
\bauthor{\bsnm{Reich},~\bfnm{Brian~J.}\binits{B.~J.}},
  \bauthor{\bsnm{Storlie},~\bfnm{Curtis~B.}\binits{C.~B.}} \AND
  \bauthor{\bsnm{Bondell},~\bfnm{Howard~D.}\binits{H.~D.}}
(\byear{2009}).
\btitle{Variable selection in {B}ayesian smoothing spline {ANOVA} models:
  Application to deterministic computer codes}.
\bjournal{Technometrics}
\bvolume{51}
\bpages{110--120}.
\bid{doi={10.1198/TECH.2009.0013}, issn={0040-1706}, mr={2668168}}
\bptok{imsref}%
\end{barticle}
\endbibitem

\bibitem[\protect\citeauthoryear{Saltelli}{2002}]{Salt2002}
\begin{barticle}[author]
\bauthor{\bsnm{Saltelli},~\bfnm{Andrea}\binits{A.}}
(\byear{2002}).
\btitle{Making best use of model evaluations to compute sensitivity indices}.
\bjournal{Comput. Phys. Comm.}
\bvolume{145}
\bpages{280--297}.
\bptok{imsref}%
\end{barticle}
\endbibitem

\bibitem[\protect\citeauthoryear{Saltelli, Chan and
  Scott}{2000}]{SaltChanScot2000}
\begin{bbook}[mr]
\beditor{\bsnm{Saltelli},~\bfnm{Andrea}\binits{A.}},
  \beditor{\bsnm{Chan.},~\bfnm{Karen}\binits{K.}} \AND
  \beditor{\bsnm{Scott},~\bfnm{E.~Marian}\binits{E.~M.}}, eds.
(\byear{2000}).
\btitle{Sensitivity Analysis}.
\bpublisher{Wiley}, \blocation{Chichester}.
\bid{mr={1886391}}
\bptok{imsref}%
\end{bbook}
\endbibitem

\bibitem[\protect\citeauthoryear{Saltelli and Tarantola}{2002}]{SaltTara2002}
\begin{barticle}[mr]
\bauthor{\bsnm{Saltelli},~\bfnm{Andrea}\binits{A.}} \AND
  \bauthor{\bsnm{Tarantola},~\bfnm{Stefano}\binits{S.}}
(\byear{2002}).
\btitle{On the relative importance of input factors in mathematical models:
  Safety assessment for nuclear waste disposal}.
\bjournal{J. Amer. Statist. Assoc.}
\bvolume{97}
\bpages{702--709}.
\bid{doi={10.1198/016214502388618447}, issn={0162-1459}, mr={1973688}}
\bptok{imsref}%
\end{barticle}
\endbibitem

\bibitem[\protect\citeauthoryear{Saltelli et~al.}{2008}]{SaltEtAl2008}
\begin{bbook}[mr]
\bauthor{\bsnm{Saltelli},~\bfnm{Andrea}\binits{A.}},
  \bauthor{\bsnm{Ratto},~\bfnm{Marco}\binits{M.}},
  \bauthor{\bsnm{Andres},~\bfnm{Terry}\binits{T.}},
  \bauthor{\bsnm{Campolongo},~\bfnm{Francesca}\binits{F.}},
  \bauthor{\bsnm{Cariboni},~\bfnm{Jessica}\binits{J.}},
  \bauthor{\bsnm{Gatelli},~\bfnm{Debora}\binits{D.}},
  \bauthor{\bsnm{Saisana},~\bfnm{Michaela}\binits{M.}} \AND
  \bauthor{\bsnm{Tarantola},~\bfnm{Stefano}\binits{S.}}
(\byear{2008}).
\btitle{Global Sensitivity Analysis. {T}he Primer}.
\bpublisher{Wiley}, \blocation{Chichester}.
\bid{mr={2382923}}
\bptok{imsref}%
\end{bbook}
\endbibitem

\bibitem[\protect\citeauthoryear{Sang and Huang}{2012}]{sanghuang2012}
\begin{barticle}[mr]
\bauthor{\bsnm{Sang},~\bfnm{Huiyan}\binits{H.}} \AND
  \bauthor{\bsnm{Huang},~\bfnm{Jianhua~Z.}\binits{J.~Z.}}
(\byear{2012}).
\btitle{A full scale approximation of covariance functions for large spatial
  data sets}.
\bjournal{J. R. Stat. Soc. Ser. B Stat. Methodol.}
\bvolume{74}
\bpages{111--132}.
\bid{doi={10.1111/j.1467-9868.2011.01007.x}, issn={1369-7412}, mr={2885842}}
\bptok{imsref}%
\end{barticle}
\endbibitem

\bibitem[\protect\citeauthoryear{Santner, Williams and
  Notz}{2003}]{santwillnotz2003}
\begin{bbook}[mr]
\bauthor{\bsnm{Santner},~\bfnm{Thomas~J.}\binits{T.~J.}},
  \bauthor{\bsnm{Williams},~\bfnm{Brian~J.}\binits{B.~J.}} \AND
  \bauthor{\bsnm{Notz},~\bfnm{William~I.}\binits{W.~I.}}
(\byear{2003}).
\btitle{The Design and Analysis of Computer Experiments}.
\bpublisher{Springer}, \blocation{New York}.
\bid{mr={2160708}}
\bptok{imsref}%
\end{bbook}
\endbibitem

\bibitem[\protect\citeauthoryear{Storlie et~al.}{2009}]{storlieetal2009}
\begin{barticle}[author]
\bauthor{\bsnm{Storlie},~\bfnm{Curtis~B.}\binits{C.~B.}},
  \bauthor{\bsnm{Swiler},~\bfnm{Laura~P.}\binits{L.~P.}},
  \bauthor{\bsnm{Helton},~\bfnm{Jon~C.}\binits{J.~C.}} \AND
  \bauthor{\bsnm{Sallaberry},~\bfnm{Cedric~J.}\binits{C.~J.}}
(\byear{2009}).
\btitle{Implementation and evaluation of nonparametric regression procedures
  for sensitivity analysis of computationally demanding models}.
\bjournal{Reliability Engineering \& System Safety}
\bvolume{94}
\bpages{1735--1763}.
\bptok{imsref}%
\end{barticle}
\endbibitem

\bibitem[\protect\citeauthoryear{Taddy, Gramacy and
  Polson}{2011}]{taddgrapols2011}
\begin{barticle}[mr]
\bauthor{\bsnm{Taddy},~\bfnm{Matthew~A.}\binits{M.~A.}},
  \bauthor{\bsnm{Gramacy},~\bfnm{Robert~B.}\binits{R.~B.}} \AND
  \bauthor{\bsnm{Polson},~\bfnm{Nicholas~G.}\binits{N.~G.}}
(\byear{2011}).
\btitle{Dynamic trees for learning and design}.
\bjournal{J. Amer. Statist. Assoc.}
\bvolume{106}
\bpages{109--123}.
\bid{doi={10.1198/jasa.2011.ap09769}, issn={0162-1459}, mr={2816706}}
\bptok{imsref}%
\end{barticle}
\endbibitem

\bibitem[\protect\citeauthoryear{Taddy et~al.}{2009}]{TaddLeeGrayGrif2009}
\begin{barticle}[mr]
\bauthor{\bsnm{Taddy},~\bfnm{Matthew~A.}\binits{M.~A.}},
  \bauthor{\bsnm{Lee},~\bfnm{Herbert K.~H.}\binits{H.~K.~H.}},
  \bauthor{\bsnm{Gray},~\bfnm{Genetha~A.}\binits{G.~A.}} \AND
  \bauthor{\bsnm{Griffin},~\bfnm{Joshua~D.}\binits{J.~D.}}
(\byear{2009}).
\btitle{Bayesian guided pattern search for robust local optimization}.
\bjournal{Technometrics}
\bvolume{51}
\bpages{389--401}.
\bid{doi={10.1198/TECH.2009.08007}, issn={0040-1706}, mr={2756475}}
\bptok{imsref}%
\end{barticle}
\endbibitem

\bibitem[\protect\citeauthoryear{Yi, Shi and Choi}{2011}]{yishichoi2011}
\begin{barticle}[mr]
\bauthor{\bsnm{Yi},~\bfnm{G.}\binits{G.}},
  \bauthor{\bsnm{Shi},~\bfnm{J.~Q.}\binits{J.~Q.}} \AND
  \bauthor{\bsnm{Choi},~\bfnm{T.}\binits{T.}}
(\byear{2011}).
\btitle{Penalized {G}aussian process regression and classification for
  high-dimensional nonlinear data}.
\bjournal{Biometrics}
\bvolume{67}
\bpages{1285--1294}.
\bid{doi={10.1111/j.1541-0420.2011.01576.x}, issn={0006-341X}, mr={2872378}}
\bptok{imsref}%
\end{barticle}
\endbibitem

\bibitem[\protect\citeauthoryear{Ziehn and Tomlin}{2009}]{ziehntomlin2009}
\begin{barticle}[author]
\bauthor{\bsnm{Ziehn},~\bfnm{T.}\binits{T.}} \AND
  \bauthor{\bsnm{Tomlin},~\bfnm{A.~S.}\binits{A.~S.}}
(\byear{2009}).
\btitle{\texttt{GUI-HDMR}---a software tool for global sensitivity analysis of
  complex models}.
\bjournal{Environmental Modelling and Software}
\bvolume{24}
\bpages{775--785}.
\bptok{imsref}%
\end{barticle}
\endbibitem

\end{thebibliography}
\end{document}